\documentclass[pra, twocolumn, floatfix, superscriptaddress]{revtex4-1}
\usepackage{amssymb,amsthm,amsmath,bbm,bbold,graphicx,epsfig,threeparttable,color}
\usepackage[colorlinks=true,linkcolor=blue]{hyperref}
\usepackage{ulem,enumerate} 
\usepackage[title]{appendix}
\usepackage{hyperref}
\usepackage{qcircuit}

\usepackage{epstopdf}

\newcommand{\ket}[1]{{|#1\rangle}}
\newcommand{\bra}[1]{{\langle#1|}}

\newcommand{\Tr}{{\rm Tr}}

\theoremstyle{definition}

\begin{document}

\title{Adaptively partitioned analog quantum simulation on near-term quantum computers: The nonclassical free-induction decay of NV centers in diamond}
\author{Yun-Hua Kuo}
\affiliation{Department of Engineering Science, National Cheng Kung University, Tainan 701401, Taiwan}
\author{Hong-Bin Chen}
\email{hongbinchen@gs.ncku.edu.tw}
\affiliation{Department of Engineering Science, National Cheng Kung University, Tainan 701401, Taiwan}
\affiliation{Center for Quantum Frontiers of Research \& Technology, NCKU, Tainan 701401, Taiwan}
\affiliation{Physics Division, National Center for Theoretical Sciences, Taipei 10617, Taiwan}

\date{\today}

\begin{abstract}
The idea of simulating quantum physics with controllable quantum devices had been proposed several decades ago. With the extensive development of quantum technology, large-scale simulation, such as
the analog quantum simulation tailoring an artificial Hamiltonian mimicking the system of interest, has been implemented on elaborate quantum experimental platforms. However, due to the limitations
caused by the significant noises and the connectivity, analog simulation is generically infeasible on near-term quantum computing platforms. Here we propose an alternative analog simulation approach
on near-term quantum devices. Our approach circumvents the limitations by adaptively partitioning the bath into several groups based on the performance of the quantum devices. We apply our approach to
simulate the free induction decay of the electron spin in a diamond NV$^-$ center coupled to a huge number of nuclei and investigate the nonclassicality induced by the nuclear spin polarization. The
simulation is implemented collaboratively with authentic devices and simulators on IBM quantum computers. We have also applied our approach to address the nonclassical noise caused by the crosstalk
between qubits. This work sheds light on a flexible approach to simulate large-scale materials on noisy near-term quantum computers.
\end{abstract}

\maketitle

\section{Introduction}

Simulating quantum physics has long been a widely known challenging problem \cite{feynman_q_sim_1982}. One of the primary difficulties lies in the exponential growth of the Hilbert space of a large
quantum system with increasing constituent components. This would require a huge amount of computer memory to store the quantum states and the quantum operations acting on them. In particular, if we are
further interested in the time evolution of the quantum system, the burden imposed on the computational resource would become even heavier and rapidly exceed the computational power of conventional
computers.

Instead of developing sophisticated, but inevitably approximate, classical algorithms, an alternative proposal for solving the problem of simulating quantum physics is to harness the power of quantum
mechanical systems \cite{feynman_q_sim_1982,barreiro_q_sim_nature_2011,georgescu_q_sim_rmp_2014,francesco_q_sim_aqt_2020,andrew_q_sim_nature_2022}, underpinned by the intuitive idea that nature itself
ultimately behaves quantum mechanically. An appealing approach is to directly map the Hamiltonian of a less controllable system of interest onto that of a quantum simulator consisting of well-controlled
quantum systems, referred to as analog quantum simulation (AQS) \cite{jae-yoon_aqs_science_2016,javier_aqs_nature_2019,jochen_aqs_nc_2017}. With the extensive development of quantum technology, AQS
has been implemented with many quantum mechanical systems, including superconducting circuit \cite{jochen_aqs_nc_2017,roushan_aqs_science_2017,jimmy_aqs_prl_2021},
ultracold atoms \cite{bloch_aqs_nat_phys_2012,javier_aqs_nature_2019}, Rydberg atoms \cite{henning_aqs_nature_2016,hannes_aqs_nature_2017},
and trapped ions \cite{islam_aqs_nc_2011,blatt_aqs_nat_phys_2012,zhang_aqs_nature_2017}. Noteworthily, these successful demonstrations of AQS are implemented on the elaborate quantum experimental
platforms, which are generically inaccessible to public.

On the other hand, many programable quantum computing platforms have emerged in recent years. They are featured by the accessibility to the public via online user interfaces, opening an avenue for the
public to experience the principles of quantum mechanics. In particular, theorists are able to design prototypical experiments running on the quantum computers to examine and demonstrate theoretical
concepts. Consequently, many demonstrations of the fundamental principles of quantum mechanics have been achieved on these state-of-the-art quantum computing platforms
\cite{bibek_qc_qi_theory_prl_2018,wright_benchmark_ionq_nc_2019,yanzhu_qc_qi_theory_pra_2019,white_qc_qi_theory_nc_2020,huanyu_qc_qi_theory_npjqi_2020,yite_qc_qi_theory_npjqi_2021,
unathi_qc_qi_theory_sr_2021,lennart_qc_qi_theory_pra_2022,fengjui_qc_qi_theory_pra_2022}.

In addition to the aforementioned demonstrations of fundamental quantum-information-theoretic principles, quantum computers are also conceived to be versatile in the simulation of open quantum system
dynamics \cite{adam_simu_open_sys_npjqi_2019,guillermo_simu_open_sys_npjqi_2020,lorenzo_simu_open_sys_prb_2020,hirsh_simu_open_sys_prxq_2022}. However, the near-term quantum computers are still in an
era of noisy intermediate-scale quantum (NISQ) devices \cite{preskill_quantum_2018}. Except for some prominent breakthroughs of quantum computers outperforming conventional computers
\cite{google_q_super_nature_2019,ibm_q_super_qst_2021,ibm_q_super_nature_2023}, due to the limitations on the performance caused by the significant noises and the qubit topological connectivity, a
straightforward simulation of large-scale materials remains intractable. Either the simulation of a few atoms arranged in an one-dimensional chain \cite{adam_simu_open_sys_npjqi_2019}, hybrid
quantum-classical algorithm \cite{bela_hyb_qc_simu_material_prx_2016,he_hyb_qc_simu_material_npjcm_2020,sirui_hyb_qc_simu_material_prxq_2021}, or variational quantum algorithms
\cite{alberto_vqa_simu_material_nc_2014,omalley_vqa_simu_material_prx_2016,cerezo_vqa_simu_material_nrp_2021}, can be efficiently implemented. There is one another approach, referred to as
Trotterization \cite{adam_simu_open_sys_npjqi_2019,lukas_trotter_npjqi_2019,markus_trotter_sci_adv_2019,myers_trotter_prr_2023}, attainable on near-term quantum computers. This approach approximates the
whole time-evolution operator by discretizing and decomposing it into a series of smaller ones according to the Suzuki-Trotter formula. The primary drawback of the Trotterization is the errors
introduced during the decomposition. Additional overhead analyzing the impacts of the decomposition errors is necessary. A quantum hardware-efficient approach capable of simulating large-scale materials
in an AQS manner free from the Trotterization decomposition errors is desirable.

On the other hand, an unambiguous demonstration of certain genuine quantumness of interest out of classicality has long been a vigorous studying topic \cite{ballentine_q_statistics_rmp_1970,
zurek_q-c_transition_rmp_2003,schlosshauer_q-c_transition_rmp_2005,modi_q-c_boundary_discord_rmp_2012}. Along with the development of quantum theory, these studies have provided deeper insights
into the quantumness of nature. Prominent paradigms includes the nonclassical correlations \cite{brunner_bell_nonlocal_rmp_2014,steering_rpp_2017,entanglement_rmp_2009} and the nonclassicality of
quantum states \cite{wigner_func_pr_1932,glauber_pr_1963,sudarshan_prl_1963}. Additionally, an emerging type of nonclassicality investigates the nature of quantum dynamical processes. Various
definitions have been put forward to elucidate different aspects of nonclassicality of quantum dynamical processes \cite{rahimikeshari_process_n_cla_prl_2013,krishna_process_n_cla_pra_2016,
jenhsiang_process_n_cla_sr_2017,smirne_process_n_cla_qst_2019,simon_kol_ext_theo_quantum_2020,simon_process_n_cla_prx_2020,alireza_process_n_cla_prl_2022,dariusz_process_n_cla_sr_2023,budini_process_n_cla_pra_2023}. Recently,
we have also approached this issue with the technique of canonical Hamiltonian ensemble representation (CHER) \cite{hongbin_process_n_cla_prl_2018,hongbin_process_n_cla_nc_2019,hongbin_cher_sr_2021} and
applied it to the free-induction-decay (FID) process of a negatively charged nitrogen-vacancy (NV$^-$) center in diamond \cite{muche_process_n_cla_nvc_jpcm_2022}.

In this work we propose an alternative analog simulation approach capable of not only simulating large-scale materials on near-term quantum computing platforms, but also reflecting the physical
mechanisms underlying the observed phenomena at a microscopic level. Our approach circumvents the limitations on the performance by adaptively partitioning the bath into several groups based on the
performance of the quantum devices. We apply our approach to simulate the FID process of the electron spin of an NV$^-$ center in diamond lattice and perform the simulation on IBM quantum computers
(IBMQ) \cite{ibmq_services}.

To do this, we first design a quantum circuit implementing the total Hamiltonian of an NV$^-$ center coupled to a huge nuclear spin bath. Additionally, to realize the effects of various nuclear spin
polarizations, we also design a family of polarization oracles accompanied with ancillary qubits. In order to adequately divide the nuclear spin bath into smaller groups fitting into the performance of
the quantum devices, we test their capabilities by a series of preliminary examinations with a few number of nuclei. Based on their performance, we can simulate the FID process either in an
collaboration with authentic quantum device and classical simulator, or fully on classical simulator of IBMQ. With this adaptive partition approach, we can reproduce the nonclassical FID process in the
presence of a transversely polarized nuclear spin bath and estimate the corresponding CHER. Noteworthily, our AQS circuit model on quantum computers is free from Trotterization decomposition errors.

To further showcase the versatility of our approach, we have also applied it to address the nonclassicality in the nonlocal noise caused by the crosstalk between qubits, which constitutes
the primary source of error in our simulations and is hard to be mitigated with post-processing of gathered data. Our approach suggests a convenient way to suppress the crosstalk by optimally grouping
the bath and launching appropriate qubits. With these paradigmatic simulation tasks, we achieve demonstration of the flexibility and capability of our approach in the exploration of new physics behind the
simulated materials.

\section{Dynamics of NV$^-$ center}\label{sec_dynamics_nvc}

Our approach is developed in the spirit of analog quantum simulation (AQS), which manipulates the tunable Hamiltonian of a well-controlled quantum system to numerically mimic a less-controllable one.
The circuit model of our appraoch will be designed specifically according to the details of the target material. Therefore, before explaining the construction of the quantum circuit,
it would be instructive to elucidate the target to be simulated.

\subsection{Hamiltonian of NV$^-$ center}

\begin{figure*}[!th]
\centering
\includegraphics[width=\textwidth]{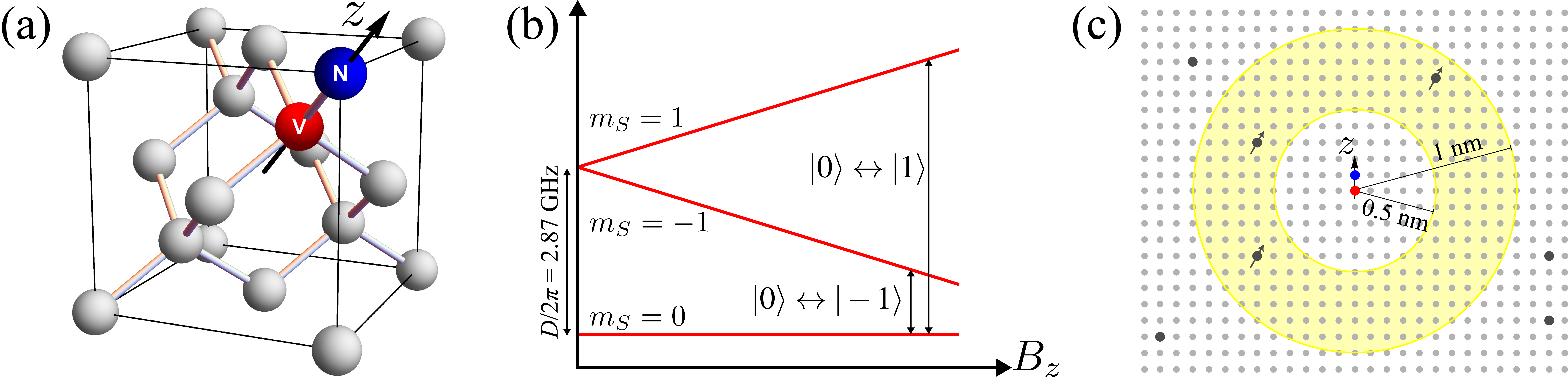}
\caption{(a) An NV$^-$ center in diamond lattice is a point defect consisting of a substitutional nitrogen (N) and a vacancy (V) in an adjacent lattice site. The axis joining V and N defines an
intrinsic $z$ axis, along which an external magnetic field $\vec{B}=B_z\vec{e}_z$ is applied. (b) For the electron spin triplet ground state, there is a zero-field splitting $D/2\pi=2.87$ GHz between
the sublevels $m_S=0$ and $m_S=\pm 1$. In the presence of an external magnetic field, the degeneracy between $m_S=\pm 1$ can be lifted due to the Zeeman splitting. Then the two different spin
transitions $\ket{0}\leftrightarrow\ket{\pm 1}$ can be selectively addressed with MW pulses at an appropriate frequency, forming a logical qubit. (c) Schematic illustration of an NV$^-$ center in diamond lattice interacting with $^{13}$C nuclear spin bath (dark gray spheres). To guarantee the validity of the dipole-dipole hyperfine interaction, all $^{13}$C nuclei lie outside a radius of 0.5 nm.
Furthermore, we also assume that only the nuclei within a polarization area (yellow shadow) of radius 1 nm can be identically polarized in a controllable manner via the DNP technique.}
\label{fig_illustration_diamond}
\end{figure*}

We consider a single negatively charged nitrogen-vacancy (NV$^-$) center in diamond lattice consisting of a substitutional nitrogen (N) and a vacancy (V) in an adjacent lattice site, as shown in
Fig.~\ref{fig_illustration_diamond}(a). The axis joining V and N defines an intrinsic $z$ axis for the electron spin. There are totally six electrons confined in the V site, forming a complicated
electron spin configuration. The ground state of the electron spin is a spin triplet state with $S=1$. Figure~\ref{fig_illustration_diamond}(b) shows the energy level structure of the electron spin
ground state. There is a zero-field splitting $D/2\pi=2.87$ GHz between the sublevels $m_S=0$ and $m_S=\pm 1$. In the absence of the external magnetic field, the two sublevels $m_S=\pm 1$ degenerate;
while the degeneracy will be lifted due to the Zeeman effect by applying an external magnetic field $\vec{B}$. For simplicity, we assume that the external magnetic field $\vec{B}=B_z\vec{e}_z$ is
aligned with the $z$ axis. Due to the Zeeman splitting, we can selectively excite the two different spin transitions $\ket{0}\leftrightarrow\ket{\pm 1}$ with microwave (MW) pulses at an appropriate
frequency. Therefore, the free Hamiltonian of the electron spin triplet is given by
\begin{equation}
\widehat{H}_\mathrm{NV}=D\widehat{S}^2_z+\gamma_eB_z\widehat{S}_z,
\label{eq_hamiltonian_nv}
\end{equation}
where $\gamma_e/2\pi=2.8025$ MHz/G is the electron gyromagnetic ratio.

The diamond lattice sites are mostly occupied by the spinless $^{12}$C nuclei [light gray spheres in Fig.~\ref{fig_illustration_diamond}(c)], which have negligible effects on the electron spin
free-induction-decay (FID) process. The electron spin dephasing is mainly caused by the randomly distributed $^{13}$C isotopes of natural abundance about 1.1$\%$ [dark gray spheres in
Fig.~\ref{fig_illustration_diamond}(c)] with nuclear spin $J=1/2$. Then the free Hamiltonian of the nuclear spin bath consisting of $^{13}$C isotopes indexed by $k$ is given by
\begin{equation}
\widehat{H}_\mathrm{C}=\sum_k\gamma_\mathrm{C}B_z\widehat{J}^{(k)}_z,
\end{equation}
with $\gamma_\mathrm{C}/2\pi=1.0704$ kHz/G being the gyromagnetic ratio of the $^{13}$C nuclei.

The coupling between the electron spin and the $^{13}$C nuclear spin bath is given by the hyperfine interaction with interaction Hamiltonian  expressed as
\begin{equation}
\widehat{H}_\mathrm{I}=\widehat{S}\cdot\sum_k\stackrel{\rightarrow}{A}^{(k)}\cdot\widehat{J}^{(k)}.
\label{eq_int_hamiltonian}
\end{equation}
Since the electron wavefunction is tightly confined in the V site, the Fermi contact risen by the overlap with the electron wavefunction becomes negligible for nuclei farther away
than 0.5 nm from the NV$^-$ center. In our simulation, we post-select a randomly generated configuration with all $^{13}$C nuclei lying outside a radius of 0.5 nm, as schematically shown in
Fig.~\ref{fig_illustration_diamond}(c). Therefore, the hyperfine interaction (\ref{eq_int_hamiltonian}) is caused by the dipole-dipole interaction and the hyperfine coefficients are given by
\begin{equation}
A_{ij}^{(k)}=\frac{\mu_0\gamma_e\gamma_\mathrm{C}}{4\pi |\vec{r}^{(k)}|^3}\left[\vec{e}_i\cdot\vec{e}_j-3(\vec{e}^{(k)}\cdot\vec{e}_i)(\vec{e}^{(k)}\cdot\vec{e}_j)\right],
\label{eq_hyperfine_coefficients}
\end{equation}
with $\mu_0$ the magnetic permeability of vacuum, $\vec{r}^{(k)}$ the displacement vector toward the $k$th nucleus, and $\vec{e}^{(k)}$ the unit vector of $\vec{r}^{(k)}$.
Note that, due to the three order of magnitude difference between $\gamma_e$ and $\gamma_\mathrm{C}$, the inter-nuclei interaction has negligible effects on the FID process.
This has also be verified with the cluster-correlation expansion technique \cite{yang_cce_prb_2009,damian_cce_njp_2021}. We therefore ignore the internuclei interaction here.

Moreover, it is worthwhile to note that the dilute $^{13}$C nuclear spin bath leads to a relaxation time $T_1$ of the electron spin in the order of milliseconds \cite{redman_nv_relaxation_prl_1991,
neumann_nv_entanglement_science_2008} and a dephasing time $T_2^\ast$ of microseconds \cite{childress_nv_center_science_2006,liu_nv_center_fid_sr_2012,maze_nv_center_fid_njp_2012}. Due to this
experimentally measured three order of magnitude difference between $T_1$ and $T_2^\ast$, the electron spin dynamics can be well approximated by pure dephasing, on the time scale under study.
Therefore, it is relevant for us to neglect the terms proportional to $\widehat{S}_x$ and $\widehat{S}_y$ in Eq.~(\ref{eq_int_hamiltonian}) and consider only the $\widehat{S}_z$ component
phenomenologically. Then the total Hamiltonian can be expressed as
\begin{equation}
\widehat{H}_\mathrm{T}=D\widehat{S}^2_z+\gamma_eB_z\widehat{S}_z+\sum_k\gamma_\mathrm{C}B_z\widehat{J}_z^{(k)}+\widehat{S}_z\sum_k\vec{A}_z^{(k)}\cdot\widehat{J}^{(k)},
\label{eq_total_hamiltonian}
\end{equation}
and only the three hyperfine coefficients $\vec{A}_z^{(k)}=(A_{zx}^{(k)},A_{zy}^{(k)},A_{zz}^{(k)})$ left. Additionally, it is critical to note that the total Hamiltonian (\ref{eq_total_hamiltonian})
can be expressed in a block diagonal form with respect to the electron spin basis as
\begin{equation}
\widehat{H}_\mathrm{T}=\sum_{m_S=0,\pm 1}\ket{m_S}\bra{m_S}\otimes\widehat{H}_{m_S},
\label{eq_total_hamiltonian_block_diag}
\end{equation}
where $\widehat{H}_{m_S}=(m_S^2D+m_S\gamma_eB_z)+\sum_k\vec{\Omega}_{m_S}^{(k)}\cdot\widehat{J}^{(k)}$, $\vec{\Omega}_{0}^{(k)}=\vec{\Omega}_0=(0,0,\gamma_\mathrm{C}B_z)$, and
$\vec{\Omega}_{\pm1}^{(k)}=\pm\vec{A}_z^{(k)}+\vec{\Omega}_0$.

Finally, the total unitary time evolution operator
\begin{eqnarray}
\widehat{U}_\mathrm{T}(t)&=&\exp(-i\widehat{H}_\mathrm{T}t) \nonumber\\
&=&\sum_{m_S=0,\pm 1}\ket{m_S}\bra{m_S}\otimes\widehat{U}_{m_S}(t),
\end{eqnarray}
is also block diagonal with respect to the electron spin basis with conditional evolution operators $\widehat{U}_{m_S}(t)=\exp(-i\widehat{H}_{m_S}t)$.

\subsection{FID process of electron spin}

The FID process of the electron spin is a pure dephasing dynamics caused by the $^{13}$C nuclear spin bath. The initial state of total system is assumed to be a direct product of
all constituent componets
\begin{equation}
\rho_\mathrm{T}(0)=\rho_\mathrm{NV}(0)\otimes\prod_k\rho^{(k)},
\label{eq_dir_prod_ini_state}
\end{equation}
where $\rho^{(k)}=[\widehat{I}^{(k)}+\vec{p}^{(k)}\cdot\hat{\sigma}^{(k)}]/2$ is the initial state of the $k$th nuclear spin with polarization $\vec{p}^{(k)}$, and $\widehat{I}^{(k)}$ and
$\hat{\sigma}^{(k)}$ are the identity and the Pauli operators, respectively, acting on the $k$th nuclear spin Hilbert space. In a conventional FID experiment, the electron spin will be first
optically polarized to $\ket{0}$ by a 532-nm green laser, and a subsequent $\pi/2$ MW pulse will set the electron spin state to a superposition state
$\ket{\Psi_\mathrm{NV}(0)}=(\ket{0}+\ket{1})/\sqrt{2}$. Therefore, in our simulation, the electron spin is described in a qubit manifold with Hilbert space spanned by the two sublevels $m_S=0$ and
$m_S=1$.

Once the electron spin state is set to $\rho_\mathrm{NV}(0)$, the hyperfine interaction in Eq.~(\ref{eq_total_hamiltonian}) is turned on and the time evolution of the total system is governed by
the block diagonal unitary operator
\begin{eqnarray}
\widehat{U}_\mathrm{T}(t)&=&\ket{0}\bra{0}\otimes\prod_k\widehat{U}_0^{(k)}(t) \nonumber\\
&&+\ket{1}\bra{1}\otimes e^{-i(D+\gamma_eB_z)t}\prod_k\widehat{U}_1^{(k)}(t),
\label{eq_total_unitary_qubit_manifold}
\end{eqnarray}
where $\widehat{U}_0^{(k)}(t)=\exp[-i(\Omega_0\hat{\sigma}^{(k)}_z)t/2]$ and $\widehat{U}_1^{(k)}(t)=\exp[-i(\vec{\Omega}_1^{(k)}\cdot\hat{\sigma}^{(k)})t/2]$, and $\vec{u}^{(k)}=\vec{\Omega}_{1}^{(k)}/|\vec{\Omega}_{1}^{(k)}|$ is the axis of nuclear spin precession.

The electron spin reduced density matrix $\rho_\mathrm{NV}(t)=\Tr_\mathrm{C}\widehat{U}_\mathrm{T}(t)\rho_\mathrm{T}(0)\widehat{U}_\mathrm{T}^\dagger(t)$ is obtained by tracing over the $^{13}$C nuclear
spin bath from the total system, and the electron spin pure dephasing dynamics is characterized by the dephasing factor
\begin{eqnarray}
\phi(t)&=&\bra{0}\rho_\mathrm{NV}(t)\ket{1} \nonumber\\
&=&e^{i(D+\gamma_eB_z)t}\prod_k\Tr\left[\widehat{U}_1^{(k)\dagger}(t)\widehat{U}_0^{(k)}(t)\rho^{(k)}\right].
\label{eq_dephasing_factor_ope_form}
\end{eqnarray}
Moreover, since we are paying particular attention to the pure dephasing dynamics caused by the $^{13}$C nuclear spin bath, it is clear that the leading factor $\exp[i(D+\gamma_eB_z)t]$ plays no role
in describing the profile of $\phi(t)$ but merely introducing a rapidly rotating phase. Consequently, for our purpose, we can neglect the leading factor. Finally, with the help of the prescription
$(\vec{u}\cdot\hat{\sigma})(\vec{v}\cdot\hat{\sigma})=(\vec{u}\cdot\vec{v})\widehat{I}+i(\vec{u}\times\vec{v})\cdot\hat{\sigma}$ and the orthogonality of the identity and the Pauli operators
$\Tr\hat{\sigma}_j\hat{\sigma}_k=2\delta_{jk}$, the dephasing factor can be expressed analytically as
\begin{eqnarray}
\phi(t)=&&\prod_k\left[\left(\cos\frac{\Omega_0t}{2}-ip_z^{(k)}\sin\frac{\Omega_0t}{2}\right)\cos\frac{\Omega_1^{(k)}t}{2} \right. \nonumber\\
&&+u_z^{(k)}\left(\sin\frac{\Omega_0t}{2}+ip_z^{(k)}\cos\frac{\Omega_0t}{2}\right)\sin\frac{\Omega_1^{(k)}t}{2} \nonumber\\
&&+i\left(p_x^{(k)}u_x^{(k)}+p_y^{(k)}u_y^{(k)}\right)\cos\frac{\Omega_0t}{2}\sin\frac{\Omega_1^{(k)}t}{2} \nonumber\\
&&\left.+i\left(p_x^{(k)}u_y^{(k)}-p_y^{(k)}u_x^{(k)}\right)\sin\frac{\Omega_0t}{2}\sin\frac{\Omega_1^{(k)}t}{2}\right].
\label{eq_dephasing_factor_use_this_form}
\end{eqnarray}

\subsection{Nuclear spin polarization}

Equation~(\ref{eq_dephasing_factor_use_this_form}) suggests that one is possible to manipulate the dynamical behavior of the electron spin by engineering the polarization $\vec{p}^{(k)}$ and the
precession axis $\vec{u}^{(k)}$ of the nuclear spin bath. One of the extensively developed techniques engineering the nuclear spin bath is the dynamical nuclear polarization (DNP)
\cite{takahashi_nuc_spin_pola_prl_2008,london_nuc_spin_pola_prl_2013,jacques_nuc_spin_pola_prl_2009,fischer_nuc_spin_pola_prl_2013,alvarez_nuc_spin_pola_nc_2015,king_nuc_spin_pola_nc_2015,
scheuer_nuc_spin_pola_njp_2016,chakraborty_nuc_spin_pola_njp_2017,scheuer_nuc_spin_pola_prb_2017,hovav_nuc_spin_pola_prl_2018,Henshaw18334_pnas_2019}, which utilizes the hyperfine interaction and the
resonance between the electron spin and the nuclei to transfer the electron spin polarization to the surrounding nuclear spins, achieving a hyperpolarized nuclear spin bath.

On the other hand, since the underlying mechanism of the DNP relies on the hyperfine interaction between the electron spin and the nuclei, which attenuates rapidly with increasing displacement, as can
be seen from Eq.~(\ref{eq_hyperfine_coefficients}), it is generically infeasible to polarize the whole nuclear spin bath. Therefore, we assume that only the nuclei within a polarization area of radius 1
nm [yellow shadow in Fig.~\ref{fig_illustration_diamond}(c)] can be polarized with identical polarization $\vec{p}$; otherwise $\vec{p}=0$ for $\vec{r}^{(k)}\geq$ 1 nm.

\section{Dynamical process nonclassicality}

From the above discussion, we acquire the fact that the decoherence of the electron spin is caused by the hyperfine interaction to the $^{13}$C nuclear spin bath. In fact, this phenomenon of decoherence
is ubiquitous in any quantum systems, as they are impossible to be fully isolated from their environments, and the inevitable interactions to their environments constitute the origin of decoherence
\cite{breuer_textbook,weiss_textbook,breuer_non_mark_review_rmp_2016,ines_non_mark_review_rmp_2017,hongbin_n_mark_pra_2017,hongbin_3sbm_scirep_2015}. From the quantum-information-theoretic perspective,
the interactions will establish complicated correlations between them; while the correlations are subject to the destructions arising from the fluctuations in the huge environments, rendering themselves
fragile and transient.

Consequently, an intriguing question is naturally raised: Given exclusively the FID signal, to what extent can the experimentalist infer the essential of the correlations between the
electron spin and the nuclear spin bath? To this end, we have developed a technique of canonical Hamiltonian ensemble representation (CHER) to characterize the nonclassicality of a dynamical process
according to the witness of the nonclassical correlations between the primary system and its environments \cite{hongbin_process_n_cla_prl_2018,hongbin_process_n_cla_nc_2019,hongbin_cher_sr_2021}.

Our definition of process nonclassicality is constructed based on the possibility to explain a dynamical process in an ensemble-averaged manner. The mathematical tool of fundamental importance in our
definition is the Hamiltonian ensemble (HE) $\{(p_{\lambda},\widehat{H}_{\lambda})\}_{\lambda}$, which consists of a collection of traceless Hermitian operators
$\widehat{H}_{\lambda}\in\mathfrak{su}(n)$ associated with a probability $p_{\lambda}$ of occurrence \cite{kropf2016effective,hongbin_disordered_sr_2022}. For a given HE, it will give rise to an
ensemble-averaged dynamics expressed as
\begin{equation}
\overline{\rho}(t)=\mathcal{E}_t\{\rho(0)\}=\int p_{\lambda}\widehat{U}_{\lambda}(t)\rho(0)\widehat{U}^{\dag}_{\lambda}(t)d\lambda,
\label{eq_ensemble-averaged-dynamics}
\end{equation}
where $\widehat{U}_{\lambda}(t) = \exp(-i\widehat{H}_{\lambda}t)$ is the unitary time-evolution operator generated by the member Hamiltonian operator $\widehat{H}_{\lambda}$.

A  particularly inspiring example considers a single qubit subject to spectral disorder with the HE given by $\{(p(\omega),\omega\hat{\sigma}_z/2)\}_\omega$, where $p(\omega)$ can be any probability
distribution function, then the ensemble-averaged dynamics describes pure dephasing:
\begin{eqnarray}
\overline{\rho}(t)&=&\int_{-\infty}^{\infty}p(\omega)e^{-i\omega\hat{\sigma}_{z}t/2}\rho_{0}\,e^{i\omega\hat{\sigma}_{z}t/2}d\omega \nonumber\\
&=&\begin{bmatrix}
\rho_{\upuparrows} & \rho_{\uparrow\downarrow}\,\phi(t) \\
\rho_{\downarrow\uparrow}\,\phi^{\ast}(t) & \rho_{\downdownarrows}
\end{bmatrix}
\label{eq_pure_dephasing}
\end{eqnarray}
with the dephasing factor $\phi(t)=\int p(\omega)\exp(-i\omega t)d\omega$ being the Fourier transform of $p(\omega)$.

Crucially, it has been shown that \cite{hongbin_process_n_cla_prl_2018}, if a primary system and its environments remain at all times classically correlated without establishing nonclassical
correlations during their interactions, then the reduced system dynamics $\mathcal{E}_t$ can be explained in terms of a HE in the sense of ensemble-averaged
dynamics~(\ref{eq_ensemble-averaged-dynamics}). Namely, the incoherent dynamical behavior can be conceived as a results of the consumption of classical correlations. On the contrary, if nonclassical
correlations emerge during the interactions, then one may fail to construct a HE with legitimate probability distribution function $p_\lambda$, and necessarily appeals to a quasi-distribution
$\wp_\lambda$ with negative values instead. Consequently, the quasi-distribution $\wp_\lambda$, referred to as the CHER, can be used to characterize the nonclassicality of a dynamics $\mathcal{E}_t$
\cite{hongbin_process_n_cla_prl_2018,hongbin_process_n_cla_nc_2019,hongbin_cher_sr_2021}.

Considering the FID process governed by the unitary operator~(\ref{eq_total_unitary_qubit_manifold}), the electron spin undergoes a pure dephasing dynamics characterized by the dephasing
factor~(\ref{eq_dephasing_factor_use_this_form}). In view of Eq.~(\ref{eq_pure_dephasing}), the corresponding CHER $\wp(\omega)$ of the electron spin FID is determined by the inverse Fourier transform
\begin{equation}
\wp(\omega)=\frac{1}{2\pi}\int_{-\infty}^\infty \phi(t) e^{i\omega t}dt.
\label{eq_inv_f_transform}
\end{equation}

It is interesting to note that the electron spin FID has shown to be nonclassical when the $^{13}$C nuclear spin bath is transversely polarized; moreover, the degree of nonclassicality will become
stronger with increasing polarization and magnetic field \cite{muche_process_n_cla_nvc_jpcm_2022}. In the following, we will design a quantum circuit capable of reproducing the nonclassical effects
induced by the nuclear spin path polarization on the electron spin FID process.

\section{Adaptively partitioned AQS for NV$^-$ center}

After elucidating the target material to be simulated and the underlying physics of nonclassicality to be revealed, we proceed to explain how to design the quantum circuit model implementing
the adaptively partitioned AQS for NV$^-$ center coupling to a huge $^{13}$C nuclear spin bath. The whole procedure consists of several steps outlined in the following:
\begin{enumerate}[\textrm{Step}~1]
\item \textbf{AQS quantum circuit}. Since the total Hamiltonian (\ref{eq_total_hamiltonian}) will generate the corresponding unitary time-evolution operator~(\ref{eq_total_unitary_qubit_manifold}), our approach begins with the design of a quantum circuit implementing Eq.~(\ref{eq_total_unitary_qubit_manifold}), as well as all the relevant experimental setup, including the initial state
    preparation and the nuclear spin polarization.
\item \textbf{Preliminary examination}. To fully simulate the effects of the whole nuclear spin bath in an AQS manner, the quantum circuit should launch several hundreds of qubits. This is obviously
    infeasible on nera-term quantum computing platforms. We therefore adaptively partition the bath into several groups based on the performance of the quantum devices. To this end, the second stage
    is to preliminarily examine the performance of available quantum devices by testing prototypical circuits designed in the first stage.
\item \textbf{Adaptive partition}. Based on the limitations of the quantum devices examined in the previous stage, the third stage is a partition of the nuclear spin bath into smaller groups fitting
    into the performances of the available quantum devices. Then each individual group is attainable on the quantum devices and can be performed separately.
\item \textbf{Combination of groups}. Ultimately, according to Eq.~(\ref{eq_dephasing_factor_use_this_form}), the final results can be obtained by combining the output of each group.
\end{enumerate}
Detailed implementations of each stage are explained in the following.



\subsection{AQS circuit model for NV$^-$ center}

\begin{figure*}[!th]
\centering
\includegraphics[width=\textwidth]{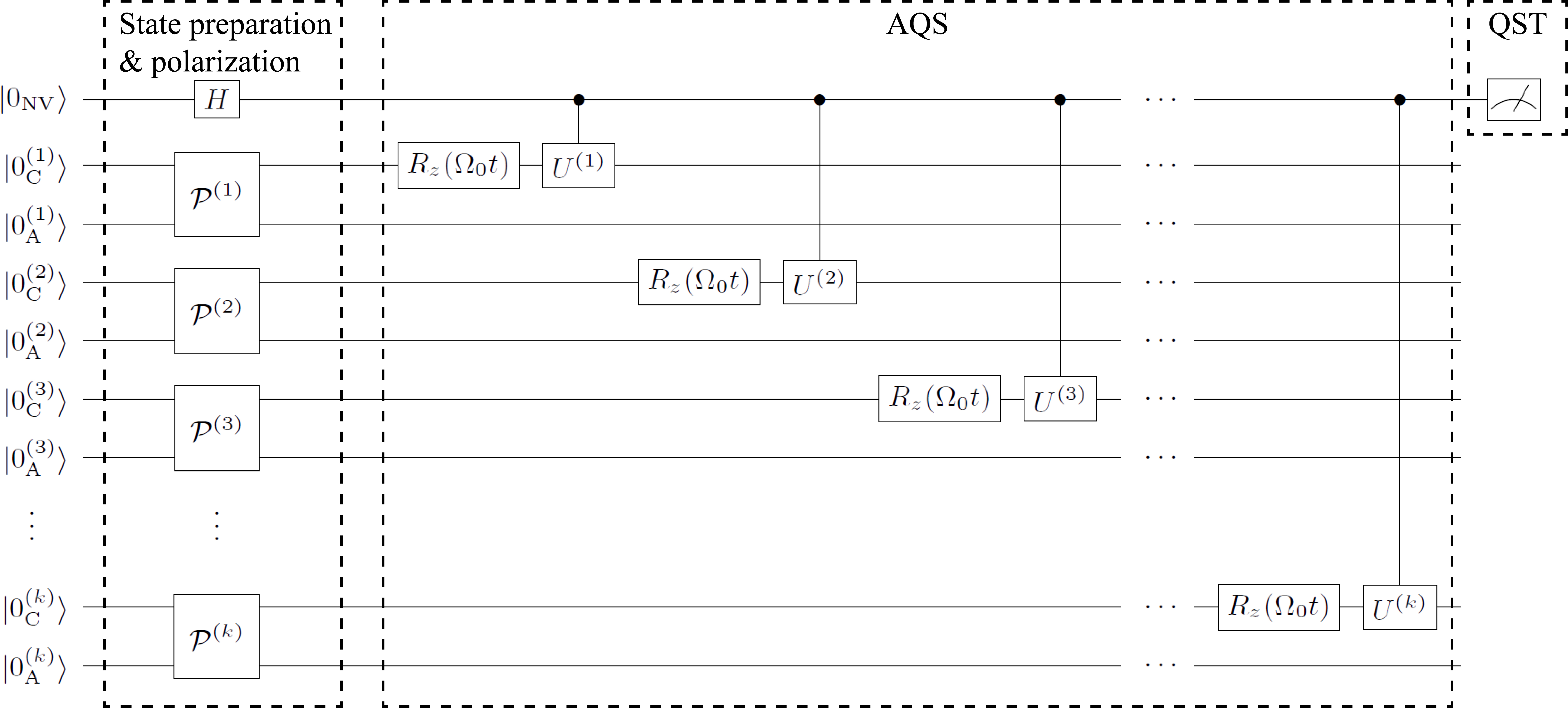}
\caption{The overall quantum circuit implementing the AQS for NV$^-$ center coupling to the whole $^{13}$C nuclear spin bath. To prepare the qubit initial states satisfying the experimental condition,
each qubit will go through a stage of state preparation. The Hadamard gate on the electron qubit sets the qubit state to $(\ket{0}+\ket{1})/\sqrt{2}$, reflecting the effect of a $\pi/2$ MW pulse. While
the mixed state of the nucleus qubit can be realized by a polarization oracle $\mathcal{P}^{(k)}$ acting on the $k$th nucleus qubit associated with an additional ancilla qubit. The desired
nuclear spin polarization can be achieved by the polarization oracles listed in Table~\ref{tab_polarization_oracle}. At the end of the electron qubit, the QST is applied to construct the time evolution
of the dephasing factor $\phi(t)$ along a time sequence.}
\label{fig_q_circuit_aqs_ibmq}
\end{figure*}

\begin{table*}
\caption{Polarization oracle and polarization vector.}
\begin{tabular}{c|ccccc}
\hline\hline
$\vec{p}^{(k)}$& $(0,0,1)$ & $(0,0,0)$ & $(0,0,\cos\theta)$ & $(1,0,0)$ & $(\sin\theta_1\sin\theta_2,0,\cos\theta_1)$ \\
\hline
 & & & &  & \\
\Qcircuit@C=0.8em @R=1.5em{
&\multigate{1}{\mathcal{P}^{(k)}}&\qw\\
&\ghost{\mathcal{P}^{(k)}}       &\qw
}                                        &
\Qcircuit@C=0.8em @R=2.4em{
&\qw     &\qw\\
&\qw     &\qw
}                                        &
\Qcircuit@C=0.8em @R=1.4em{
&\gate{H}&\ctrl{1}&\qw\\
&\qw     &\targ   &\qw
}                                        &
\Qcircuit@C=0.8em @R=1.2em{
&\gate{U(\theta,0,0)}&\ctrl{1}&\qw\\
&\qw                 &\targ   &\qw
}                                        &
\Qcircuit@C=0.8em @R=1.6em{
&&\gate{U(\pi/2,0,0)}&\qw\\
&&\qw                &\qw
}                                        &
\Qcircuit@C=0.8em @R=0.8em{
&&\gate{U(\theta_1,0,0)}&\ctrl{1}&\qw\\
&&\gate{U(\theta_2,0,0)}&\targ   &\qw
}                                        \\
 & & & &  & \\
\hline\hline
\end{tabular}
\label{tab_polarization_oracle}
\end{table*}

The purpose of the AQS is to
tailor an artificial Hamiltonian with controllable quantum systems mimicking the one of interest. We therefore design a quantum circuit by mapping the total unitary time-evolution operator
\begin{equation}
\widehat{U}_\mathrm{T}(t)=\ket{0}\bra{0}\otimes\prod_k\widehat{U}_0^{(k)}(t)+\ket{1}\bra{1}\otimes\prod_k\widehat{U}_1^{(k)}(t)
\label{eq_total_unitary_use_this_form}
\end{equation}
into quantum gates. Note that the factor $\exp[-i(D+\gamma_eB_z)t]$ has been neglected from Eq.~(\ref{eq_total_unitary_qubit_manifold}). This factor is given by the energy-level spacing between
$\ket{0}$ and $\ket{1}$ states described by the electron spin free Hamiltonian~(\ref{eq_hamiltonian_nv}). It is responsible for the rapid oscillation in the FID profile. However, here we are interested
in the dephasing caused by the interaction to the nuclear spin bath. Consequently, for our purpose, we can neglect the this factor.

It is crucial to observe that the hyperfine interaction in Eq.~(\ref{eq_total_hamiltonian}) gives rise to an intrinsic conditional operation conditioned on the electron spin state. This can be realized
by the controlled-U gates on IBMQ after the following manipulation of Eq.~(\ref{eq_total_unitary_use_this_form}):
\begin{eqnarray}
\widehat{U}_\mathrm{T}(t)&=&\left(\ket{0}\bra{0}\otimes\prod_k\widehat{I}^{(k)}+\ket{1}\bra{1}\otimes\prod_k\widehat{U}_1^{(k)}(t)\widehat{U}_0^{(k)\dagger}(t)\right) \nonumber\\
&&\times\left(\widehat{I}^{(\mathrm{NV})}\otimes\prod_k\widehat{U}_0^{(k)}(t)\right),
\end{eqnarray}
where $\widehat{I}^{(\mathrm{NV})}$ is the identity operator acting on the qubit playing the role of electron spin. Then the above unitary operator can be realized with quantum gates as:
\begin{eqnarray}
\widehat{U}_\mathrm{T}&=&\left(\mathrm{C}^{(\mathrm{NV})}\prod_k\widehat{U}^{(k)}(\theta^{(k)},\varphi^{(k)},\lambda^{(k)},\gamma^{(k)})\right) \nonumber\\
&&\times\left(\widehat{I}^{(\mathrm{NV})}\otimes\prod_k\widehat{R}_{z}(\Omega_0t)\right).
\label{eq_total_unitary_qc_implementation}
\end{eqnarray}
The second term denotes a series of identical and independent $\widehat{R}_{z}(\Omega_0t)$ rotations, with matrix representation
\begin{equation}
\Qcircuit @C=0.8em @R=0.75em {
	& \gate{R_{z}(\theta)} & \qw
}	= e^{-i\frac{\theta}{2}\hat{\sigma}^{(k)}_z} =
\left[\begin{array}{cc}
e^{-i\frac{\theta}{2}} & 0 \\
0 & e^{i\frac{\theta}{2}}
\end{array} \right],
\end{equation}
on the qubits playing the role of $^{13}$C nuclear spins, followed by the controlled-U gates conditioned on the electron qubit denoted by the first term. They can be realized by the circuit
\begin{equation}
\begin{array}{c}
\Qcircuit@C=0.8em @R=0.75em{
&\ctrl{2}      &\qw&                                          &&\ctrl{2}&\qw \\
&              &   &\push{\rule{0.1em}{0em}=\rule{0.1em}{0em}}&&     \\
&\gate{U^{(k)}}&\qw&                                          &&\gate{e^{i\gamma^{(k)}}U(\theta^{(k)},\varphi^{(k)},\lambda^{(k)})} & \qw
}
\end{array}\end{equation}
on IBMQ; meanwhile, the gate parameters can be determined according to the Hamiltonian~(\ref{eq_total_hamiltonian_block_diag}) as follows:
\begin{equation}
\left\{\begin{array}{l}
\theta^{(k)}=2\cos^{-1}\sqrt{\cos^2\frac{\Omega^{(k)}_1 t}{2}+\sin^2\frac{\Omega^{(k)}_1 t}{2}u_z^{(k)2}} \\
\varphi^{(k)}=-\frac{\pi}{2}-\Theta^{(k)}+\Phi^{(k)}\\
\lambda^{(k)}=\frac{\pi}{2}-\Omega_0t-\Theta^{(k)}-\Phi^{(k)}\\
\gamma^{(k)}=\frac{\Omega_0t}{2}+\Theta^{(k)}\\
\Theta^{(k)}=\mathrm{Arg}\left[\cos\frac{\Omega^{(k)}_1 t}{2}-i\sin\frac{\Omega^{(k)}_1 t}{2}u^{(k)}_z\right]\\
\Phi^{(k)}=\mathrm{Arg}\left[u^{(k)}_x + iu^{(k)}_y\right]
\end{array}\right..
\label{eq_parameter_cu_aqs}
\end{equation}
Further details are shown in Appendix~\ref{app_parameter_cu_aqs}. Consequently, the total unitary time-evolution operator~(\ref{eq_total_unitary_use_this_form}) can be realized with the AQS
circuit succinctly shown below:
\begin{equation}
\begin{array}{c}
\Qcircuit@C=0.8em @R=0.75em{
&      &                              &   &                                          &&      &\push{\mathrm{AQS}}     &\\
&\qw   &\multigate{2}{U_\mathrm{T}(t)}&\qw&                                          &&\qw   &\qw                     &\ctrl{2}      &\qw\\
&      &                              &   &\push{\rule{0.1em}{0em}=\rule{0.1em}{0em}}&\\
&{/}\qw&\ghost{U_\mathrm{T}(t)}       &\qw&                                          &&{/}\qw&\gate{R_{z}(\Omega_0 t)}&\gate{U^{(k)}}&\qw
}
\end{array}.\end{equation}

\subsection{State preparation and polarization oracle}

Once the total unitary time-evolution operator~(\ref{eq_total_unitary_use_this_form}) has been realized with quantum circuit, following the discussions in Sec.~\ref{sec_dynamics_nvc}, the next step is
to prepare the initial state as given in Eq.~(\ref{eq_dir_prod_ini_state}) according to the FID experiments.

The qubit initial state on IBMQ is preset to $\ket{0}$. An Hadamard gate realizes the effect of a $\pi/2$ MW pulse setting the electron spin state to $(\ket{0}+\ket{1})/\sqrt{2}$, as shown in
Fig.~\ref{fig_q_circuit_aqs_ibmq}. On the other hand, a single-qubit gate on nucleus qubit is insufficient to realize various nuclear spin states, particularly those of mixed states. To do this, we
design the polarization oracle $\mathcal{P}^{(k)}$ acting on the $k$th nucleus qubit associated with an additional ancilla qubit, as shown in Fig.~\ref{fig_q_circuit_aqs_ibmq}. After the operation of an appropriate $\mathcal{P}^{(k)}$, tracing out the ancilla qubit leaves the nucleus qubit in the state $\rho^{(k)}=[\widehat{I}^{(k)}+\vec{p}^{(k)}\cdot\hat{\sigma}^{(k)}]/2$ with a
corresponding polarization vector $\vec{p}^{(k)}$. Table~\ref{tab_polarization_oracle} shows a family of polarization oracles $\mathcal{P}^{(k)}$ and the corresponding polarization vectors
$\vec{p}^{(k)}$. Therefore, we can manipulate individual nucleus qubit state and realize a nuclear spin bath of experimental condition schematically shown in Fig.~\ref{fig_illustration_diamond}(c).

At the end of the AQS circuit, the quantum state tomography (QST) is applied to probe the state of the electron qubit. Additionally, since we are aiming at simulating the electron spin pure dephasing
characterized by the dephasing factor~(\ref{eq_dephasing_factor_use_this_form}), its time evolution can be constructed by measuring $\hat{\sigma}_x$ and $\hat{\sigma}_y$ along a time sequence according
to $\phi(t)=\langle\hat{\sigma}_x\rangle_t-i\langle\hat{\sigma}_y\rangle_t$. Finally, the overall layout of the circuit is shown in Fig.~\ref{fig_q_circuit_aqs_ibmq}. Note that merely the electron qubit
is measured for QST at the end of the AQS circuit. The nucleus and the ancilla qubits are ignored after the AQS block. This reflects the trace over the nuclear spin degrees of freedom in
Eq.~(\ref{eq_dephasing_factor_ope_form}).

\subsection{Preliminary examination} \label{sec_preliminary_exam}

\begin{figure}[!th]
\centering
\includegraphics[width=\columnwidth]{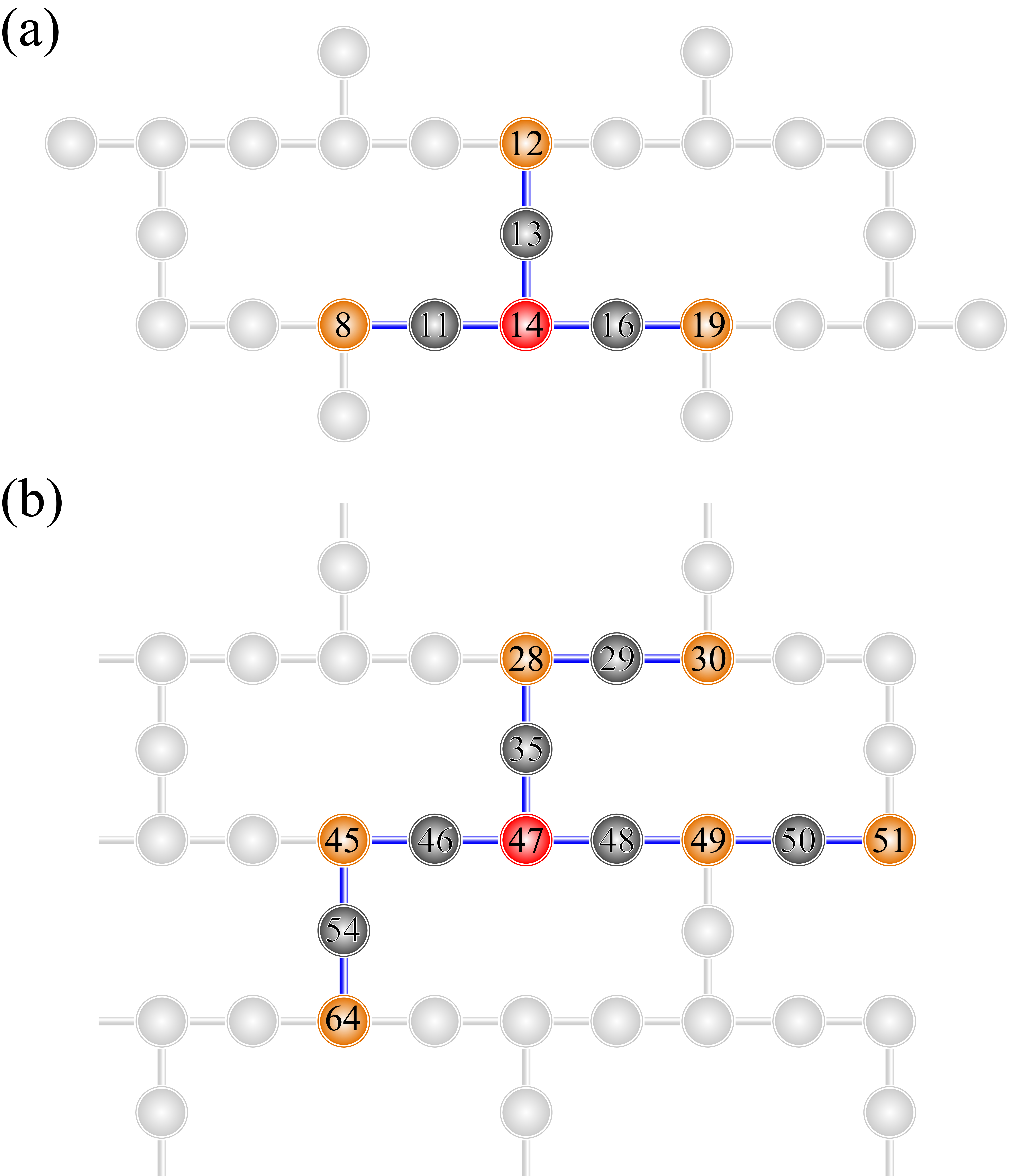}
\caption{The qubits launched in the simulation on the (a) \texttt{ibm\_auckland} and (b) \texttt{ibm\_washington} quantum devices. The red, dark gray, and orange qubits play the role of the electron
spins, the nuclear spins, and the ancilla qubits controlling the nuclear spin polarizations, respectively.}
\label{fig_layout_ibmq_in_use}
\end{figure}

\begin{figure*}[th]
\centering
\includegraphics[width=\textwidth]{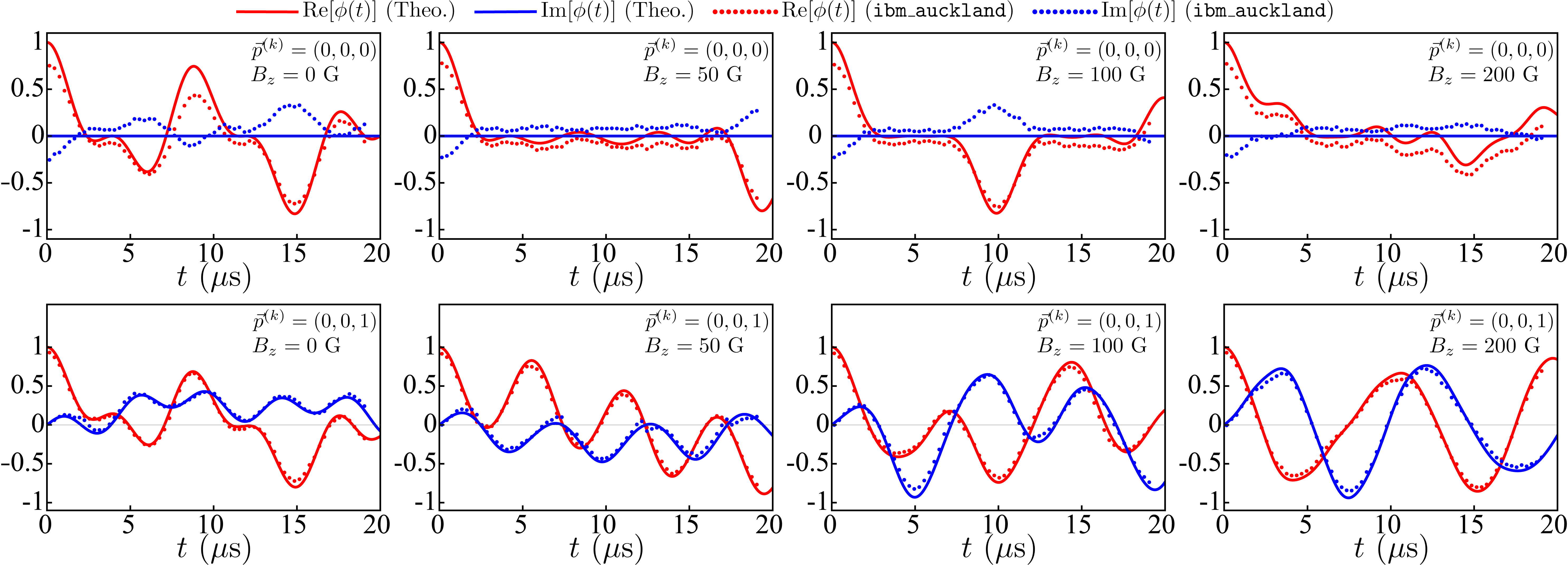}
\caption{The AQS results for three nuclei obtained from \texttt{ibm\_auckland}. We demonstrate the results for two polarizations, $\vec{p}^{(k)}=(0,0,0)$ (top panels) and $\vec{p}^{(k)}=(0,0,1)$
(bottom panels), at various values of the magnetic field. The simulation results for $\vec{p}^{(k)}=(0,0,1)$ fit the theoretical calculations well since the polarization corresponds to the preset qubit
state $\ket{0}$ without additional operation. On the other hand, to prepare the nuclear spin polarization $\vec{p}^{(k)}=(0,0,0)$ requires a CNOT gate, resulting in obvious discrepancies, particularly
the erroneous imaginary part $\mathrm{Im}[\phi(t)]$.}
\label{fig_3c_total}
\end{figure*}

\begin{figure*}[th]
\centering
\includegraphics[width=\textwidth]{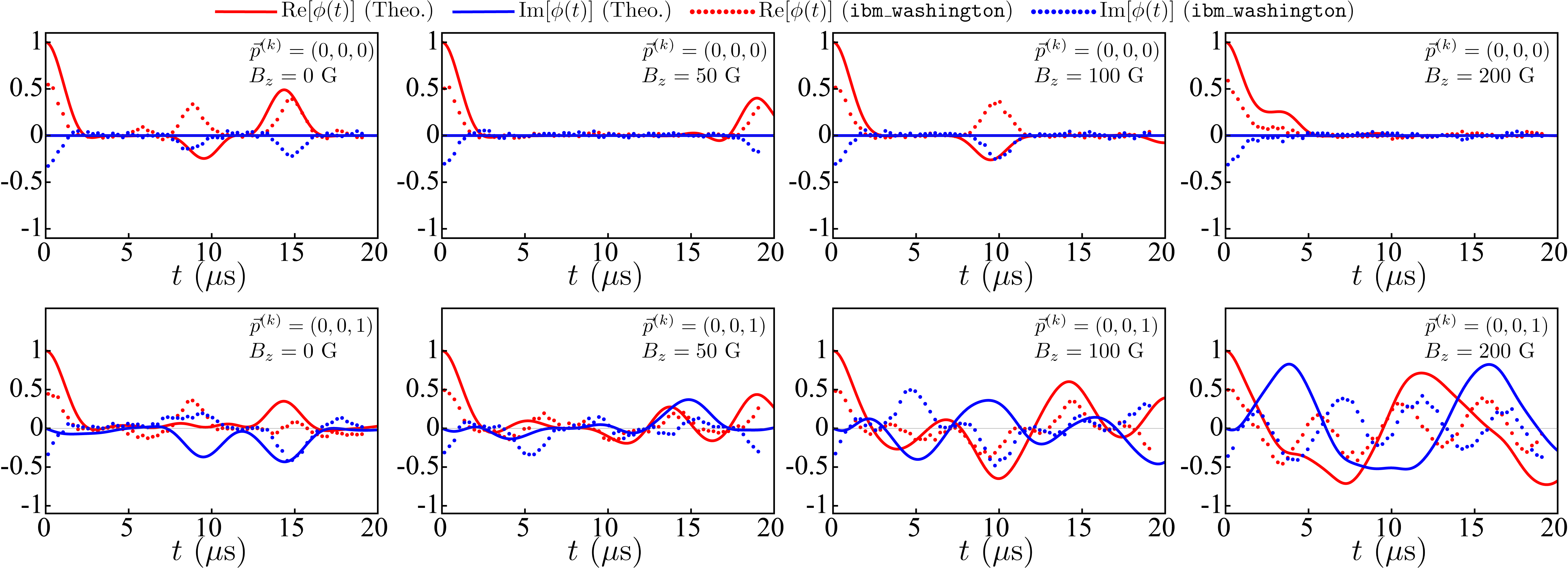}
\caption{The AQS results for six nuclei obtained from \texttt{ibm\_washington}. We demonstrate the results for two polarizations, $\vec{p}^{(k)}=(0,0,0)$ (top panels) and $\vec{p}^{(k)}=(0,0,1)$
(bottom panels), at various values of the magnetic field. Due to the limitation imposed by the topological connectivity of IBMQ devices, nucleus qubits exceeding three will lie at farther positions away
from the electronic qubit, resulting in a rapidly increasing number of CNOT gates. This not only enhances the noise, but also deepens the circuit, rendering the simulation unreliable.}
\label{fig_6c_total}
\end{figure*}

To perform the AQS circuit on IBMQ \cite{ibmq_services}, we have to map the circuit onto the qubits of the quantum devices. However, due to the qubit topological connectivity, it is obviously infeasible
to map the whole circuit simulating hundreds of nucleus qubits onto IBMQ devices.

To verify the validity of the circuit, as well as to benchmark the performance of the IBMQ devices for later purpose, we first perform two prototypical circuits simulating the effects of three and six
$^{13}$C nuclei on \texttt{ibm\_auckland} and \texttt{ibm\_washington}, respectively. The qubits launched and the labels on IBMQ devices are shown in Fig.~\ref{fig_layout_ibmq_in_use}. The red qubits
play the role of the electron spin, and the dark gray and orange qubits denote the nucleus and the ancilla qubits controlling the nuclear spin polarizations, respectively.

Figures~\ref{fig_3c_total} and \ref{fig_6c_total} show the results of the prototypical simulations of three and six nuclei, respectively. We demonstrate the results of two polarizations, i.e.,
$\vec{p}^{(k)}=(0,0,0)$ and $(0,0,1)$, at various values of the magnetic field. In Fig.~\ref{fig_3c_total}, the results obtained from \texttt{ibm\_auckland} for $\vec{p}^{(k)}=(0,0,1)$ (bottom panels)
are in good agreement with the theoretical calculations given by Eq.~(\ref{eq_dephasing_factor_use_this_form}); while the ones for $\vec{p}^{(k)}=(0,0,0)$ (top panels) show prominent
discrepancies. These discrepancies can be understood from two aspects. The first one is the polarization oracles listed in Table~\ref{tab_polarization_oracle}. Polarization $\vec{p}^{(k)}=(0,0,1)$
corresponds to the preset qubit state $\ket{0}$ without additional operation. However, the one for $\vec{p}^{(k)}=(0,0,0)$ requires an additional CNOT gate coupling to an ancilla qubit for each nucleus
qubit, which constitutes one source of the noise on IBMQ devices. Later we will further investigate the second source of the nonlocal noise caused by the crosstalk between qubits on IBMQ devices. We will find that this nonlocal noise is also nonclassical, and constitutes the primary source of error, particularly the erroneous imaginary part $\mathrm{Im}[\phi(t)]$.

Additionally, we have also increased the number of nuclei to six and shown the results in Fig.~\ref{fig_6c_total}. It can be seen that the results obtained from \texttt{ibm\_washington} deviate even
more considerably from the theoretical calculations. The reason for this enhanced deviation can be understood from the topological connectivity of IBMQ devices. As shown in
Fig.~\ref{fig_layout_ibmq_in_use}(b), an electron qubit can at most physically connect to three nucleus qubits, to each of which an additional ancilla qubit is appended. Further nucleus
qubits will lie at farther positions away from the electronic qubit, leading to remotely controlled-U gates. Due to the limited connectivity, the remotely controlled-U gates are implemented in the back
end by appending additional SWAP gates as
\begin{equation}
\begin{array}{c}
\Qcircuit@C=0.8em @R=0.75em{
&\ctrl{3}      &\qw&                                          &&\qswap    &\qw       &\qw           &\qw       &\qswap    &\qw \\
&\qw           &\qw&                                          &&\qswap\qwx&\qswap    &\qw           &\qswap    &\qswap\qwx&\qw \\
&\qw           &\qw&\push{\rule{0.1em}{0em}=\rule{0.1em}{0em}}&&\qw       &\qswap\qwx&\ctrl{1}      &\qswap\qwx&\qw       &\qw \\
&\gate{U^{(k)}}&\qw&                                          &&\qw       &\qw       &\gate{U^{(k)}}&\qw       &\qw       &\qw
}
\end{array};\end{equation}
and each swap gate will introduce three more CNOT gates as
\begin{equation}
\begin{array}{c}
\Qcircuit@C=0.8em @R=0.75em{
&\qswap     &\qw&                                          &&\ctrl{2}&\targ    &\ctrl{2}&\qw \\
&\qwx       &   &\push{\rule{0.1em}{0em}=\rule{0.1em}{0em}}&&\\
&\qswap\qwx &\qw&                                          &&\targ   &\ctrl{-2}&\targ   &\qw
}
\end{array}.\end{equation}
This results in a rapidly increasing number of CNOT gates in the back end implementation, as well as the detrimental noises. Furthermore, an increasing number of CNOT gates also implies a deeper circuit,
which requires a longer execution time approaching, or even exceeding, the life time of physical qubits, rendering the results unreliable.

Finally, we have also performed the AQS for ten nuclei on \texttt{ibmq\_qasm\_simulator}. We find that the results from the simulator fit the theoretical calculations very well besides tiny
errors due to the approximations introduced by classical simulation algorithm; whereas, this simulator has a limited computational capability and can simulate the effects of at most ten
nucleus-ancilla qubit pairs in a single task. The results and further discussions are shown in Appendix~\ref{simulate 10 nuclei on simulator}.

\subsection{Adaptive partition of the bath}

\begin{table*}
\caption{Partition of the 520 nuclei and the implementation of each group.}
\begin{tabular}{c|ccccccccc}
\hline\hline
$k$                   & $|\vec{r}^{(k)}|$ (nm)&$\quad$& $\vec{p}^{(k)}$ &$\quad$& Device                         &$\quad$& Amount of $^{13}$C  &$\quad$& Amount of qubits \\
\hline
$\sharp$1-$\sharp$3   & $0.5 \sim 1$          &$\quad$& Controllable    &$\quad$& \texttt{ibm\_auckland}         &$\quad$& 3                   &$\quad$& 7 \\
$\sharp$4-$\sharp$6   & $0.5 \sim 1$          &$\quad$& Controllable    &$\quad$& \texttt{ibm\_auckland}         &$\quad$& 3                   &$\quad$& 7 \\
$\sharp$7-$\sharp$9   & $0.5 \sim 1$          &$\quad$& Controllable    &$\quad$& \texttt{ibm\_auckland}         &$\quad$& 3                   &$\quad$& 7 \\
$\sharp$10            & $0.5 \sim 1$          &$\quad$& Controllable    &$\quad$& \texttt{ibm\_auckland}         &$\quad$& 1                   &$\quad$& 3 \\
\hline
$\sharp$11-$\sharp$20 & $>1$                  &$\quad$& $(0,0,0)$       &$\quad$& \texttt{ibmq\_qasm\_simulator} &$\quad$& 10                  &$\quad$& 21 \\
$\sharp$21-$\sharp$30 & $>1$                  &$\quad$& $(0,0,0)$       &$\quad$& \texttt{ibmq\_qasm\_simulator} &$\quad$& 10                  &$\quad$& 21 \\
$\vdots$              &                       &$\quad$&                 &$\quad$&                                &$\quad$&                     &$\quad$&    \\
$\sharp$511-$\sharp$520& $>1$                 &$\quad$& $(0,0,0)$       &$\quad$& \texttt{ibmq\_qasm\_simulator} &$\quad$& 10                  &$\quad$& 21 \\
\hline\hline
\end{tabular}
\label{tab_partition}
\end{table*}

From the previous preliminary examinations, it can be seen that the number of nuclei simulated in a single task is very limited, far from simulating large-scale materials in an AQS manner. To circumvent
these limitations, we design a simulation algorithm by adaptively dividing the nuclear spin bath into several groups, each of which fits into the performance of the
quantum devices.

In our simulation, we first generate a nuclear spin configuration of natural abundance about 1.1$\%$, consisting of 520 $^{13}$C nuclei randomly distributed over the diamond lattice sites.
Then we list the nuclei according to the distance $|\vec{r}^{(k)}|$ to the electron spin in an increasing order.
To ensure the validity of the dipole-dipole interaction described by Eq.~(\ref{eq_hyperfine_coefficients}), we have also verified that all nuclei are farther away than 0.5 nm from the electron spin.

Table~\ref{tab_partition} shows how we partition the 520 nuclei. For example, the first row denotes a group consisting of three nuclei lying within the polarization area
($0.5~\mathrm{nm}<|\vec{r}^{(k)}|<1~\mathrm{nm}$). Then the effect can be simulated on \texttt{ibm\_auckland} with a circuit launching seven qubits [Fig.~\ref{fig_layout_ibmq_in_use}(a)], and the
polarization vector $\vec{p}^{(k)}$ is controllable with appropriate polarization oracle listed in Table~\ref{tab_polarization_oracle}. In our configuration, there are ten nuclei lying within the
polarization area. For the unpolarized nuclei with $\vec{p}^{(k)}=(0,0,0)$ outside the polarization area, e.g., the group consisting of nuclei ranging from $\sharp$11 to $\sharp$20, the circuits are
simulated on \texttt{ibmq\_qasm\_simulator}.

Then the effects of the whole nuclear spin bath are implemented in a collaboration between the authentic device \texttt{ibm\_auckland} and the simulator \texttt{ibmq\_qasm\_simulator} on IBMQ.
Finally, according to Eq.~(\ref{eq_dephasing_factor_use_this_form}), the desired dephasing factor $\phi(t)$ accounting for 520 nuclei is given by the product of the results of all groups, and the
corresponding CHER $\wp(\omega)$ can be estimated according to the inverse Fourier transform~(\ref{eq_inv_f_transform}).

\section{Simulation results}

\begin{figure*}[th]
\centering
\includegraphics[width=\textwidth]{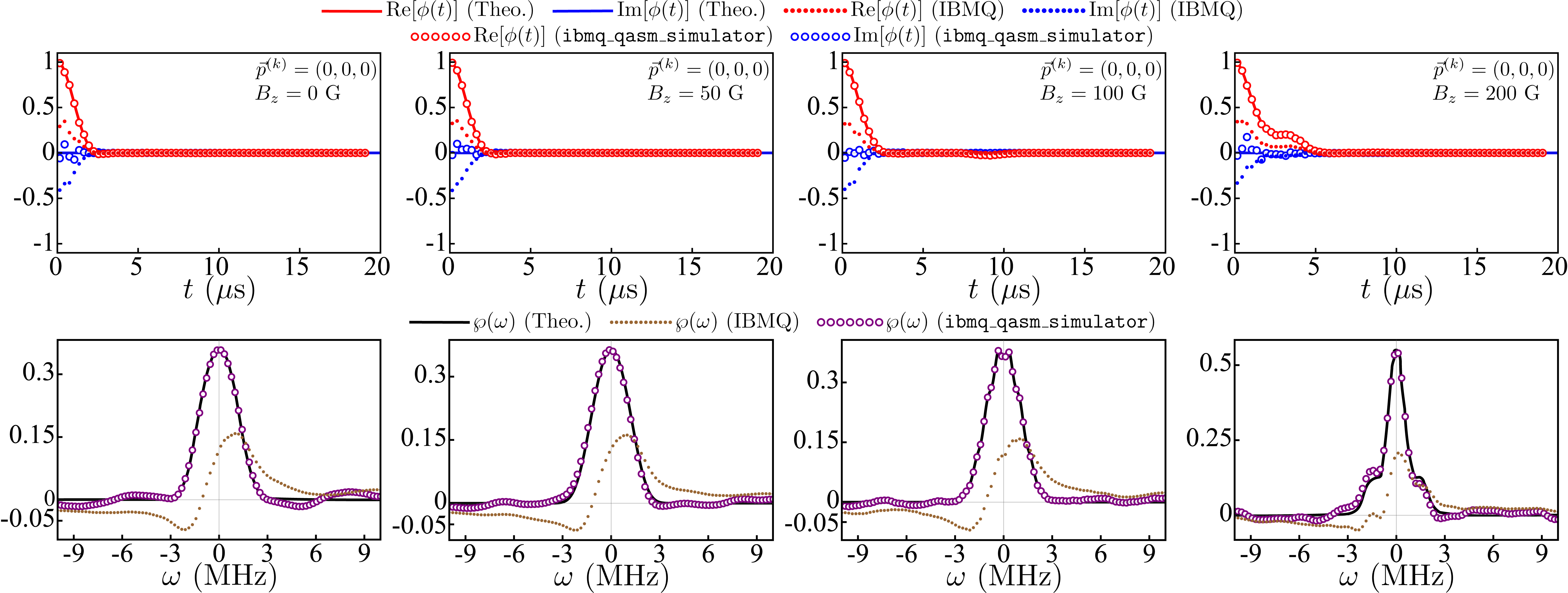}
\caption{The adaptively partitioned AQS results for 520 unpolarized nuclei at various values of the magnetic field (top panels) and the corresponding CHER (bottom panels). The collaborative simulations
with \texttt{ibm\_auckland} and \texttt{ibmq\_qasm\_simulator} are denoted by the colored dots, and the ones fully given by \texttt{ibmq\_qasm\_simulator} are denoted by the colored circles. The errors
caused by the \texttt{ibm\_auckland} are prominent, particularly in the beginning of the time evolution. The simulator gives better results besides the amplified algorithmic errors in the imaginary
part. Although the CHER in the case of unpolarized nuclear spin bath should be positive, the errors caused by the crosstalk on \texttt{ibm\_auckland} give rise to negative wings. On the other hand, the
results fully given by \texttt{ibmq\_qasm\_simulator} reproduce the central peak very well; while the algorithmic errors give rise to irregularly wavy wings on both sides of the central peak.}
\label{fig_520c_000_Fourier_total}
\end{figure*}

\begin{figure*}[th]
\centering
\includegraphics[width=\textwidth]{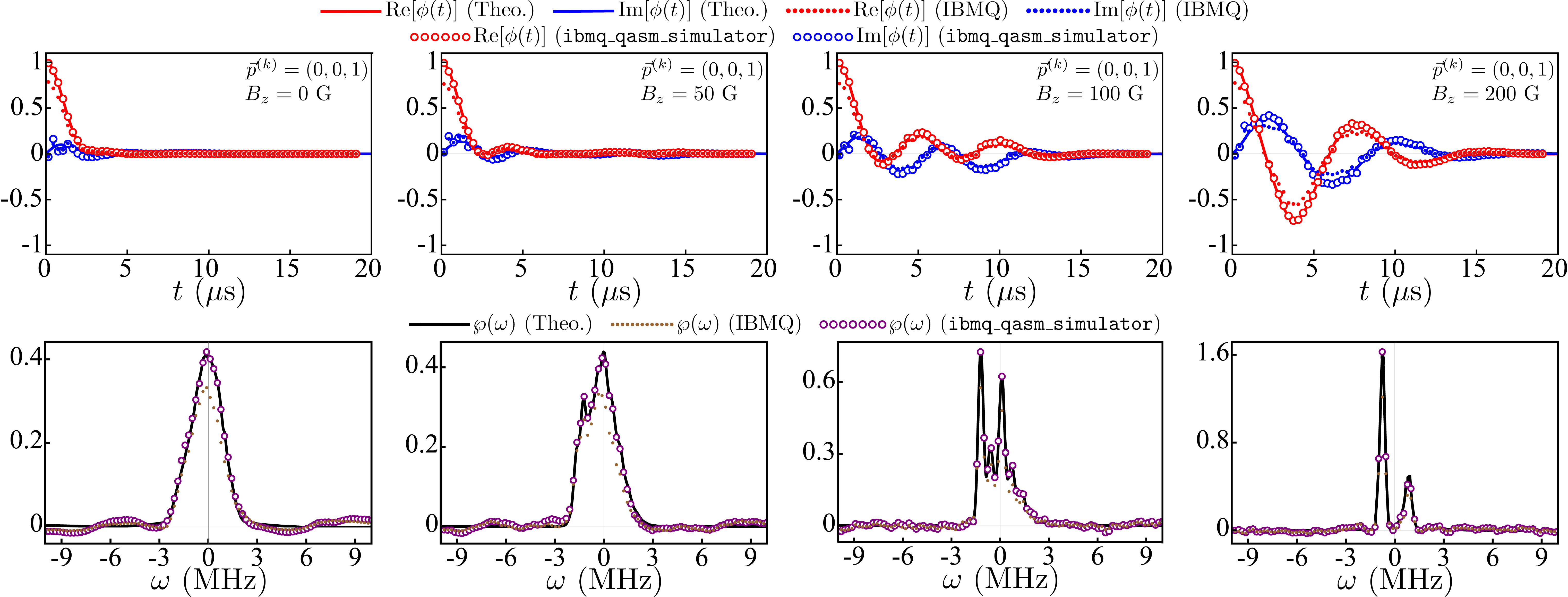}
\caption{The adaptively partitioned AQS results for a $z$-polarized nuclear spin bath at various values of the magnetic field (top panels) and the corresponding CHER (bottom panels). The collaborative
simulations with \texttt{ibm\_auckland} and \texttt{ibmq\_qasm\_simulator} are denoted by the colored dots, and the ones fully given by \texttt{ibmq\_qasm\_simulator} are denoted by the colored
circles. Due to the null operation of the polarization oracle implementing $\vec{p}^{(k)}=(0,0,1)$, the collaborative simulations are in good agreement with the theoretical calculations. Remarkably, the
emergence of the sharp peaks in the profile of the CHER has also been well-reproduced in our simulations.}
\label{fig_520c_001_Fourier_total}
\end{figure*}

\begin{figure*}[th]
\centering
\includegraphics[width=\textwidth]{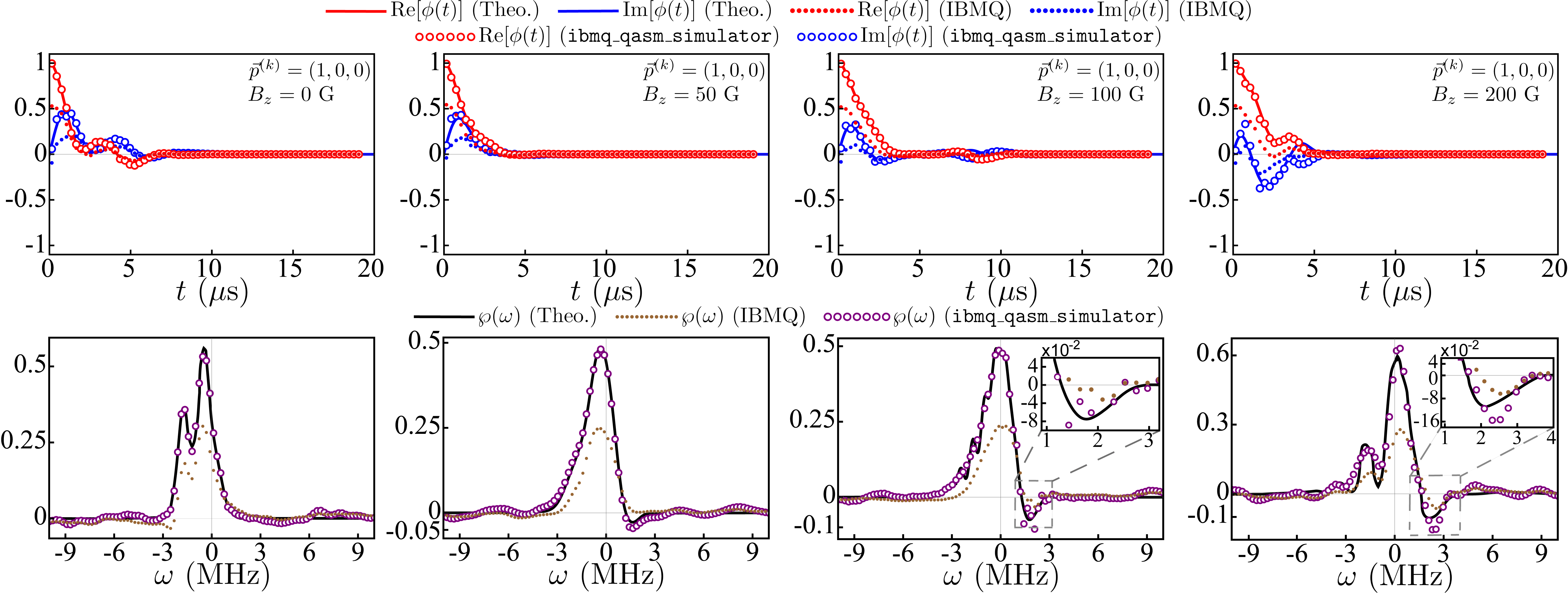}
\caption{The adaptively partitioned AQS results for an $x$-polarized nuclear spin bath at various values of the magnetic field (top panels) and the corresponding CHER (bottom panels). The collaborative
simulations with \texttt{ibm\_auckland} and \texttt{ibmq\_qasm\_simulator} are denoted by the colored dots, and the ones fully given by \texttt{ibmq\_qasm\_simulator} are denoted by the colored
circles. Due to the polarization oracle implementing $\vec{p}^{(k)}=(1,0,0)$ on \texttt{ibm\_auckland}, the collaborative simulations ultimately deviate prominently from the theoretical calculations.
In this case, the ones fully given by \texttt{ibmq\_qasm\_simulator} also suffer from the amplified algorithmic errors. Remarkably, the negativity in the CHER $\wp(\omega)$ is enhanced against the
errors at stronger fields and becomes visible, as shown in the insets. This is an indicator of the nonclassicality reproduced in our simulations.}
\label{fig_520c_100_Fourier_total}
\end{figure*}

We first show the results in Fig.~\ref{fig_520c_000_Fourier_total} for an unpolarized nuclear spin bath, i.e., $\vec{p}^{(k)}=(0,0,0)$ for both the ten nuclei simulated on \texttt{ibm\_auckland}
and the outer nuclei on \texttt{ibmq\_qasm\_simulator}, denoted by the colored dots. As expected from the top panels of Fig.~\ref{fig_3c_total}, we can observe significant errors in
Fig.~\ref{fig_520c_000_Fourier_total}, particularly in the beginning of the time evolution. As a comparative study, we also demonstrate a counterpart fully performed on \texttt{ibmq\_qasm\_simulator},
denoted by the colored circles. Although the simulator gives better results than those of collaborative simulation, the algorithmic errors now become visible in the imaginary parts, due to the
amplification caused by the production over all groups of nuclei.

In the bottom panels of Fig.~\ref{fig_520c_000_Fourier_total}, we show the corresponding CHER $\wp(\omega)$ at various values of the magnetic field. The theoretical calculations show that the CHER should
be positive in the case of unpolarized nuclear spin bath, whereas the errors caused by \texttt{ibm\_auckland} result in negative wings. In view of the physical meaning of the negativity as a witness of
nonclassical system-environment correlations \cite{hongbin_process_n_cla_prl_2018}, the negative wings imply that there are certain nonclassical, and nonlocal, correlations established between the
nucleus-ancilla superconducting devices and the environmental degrees of freedom in the substrate during the pulse operations. This effect is referred to as the crosstalk between the nucleus-ancilla
qubit pairs and gives rise to nonlocal noises between qubit pairs, which in turn come into play in the dynamics of the electron spin qubit and is captured by the negativity in the CHERs. Later we will
address this issue by suppressing its effect with appropriate qubit pairs. On the other hand, the results fully given by \texttt{ibmq\_qasm\_simulator} reproduce the central peak very well. However, the
algorithmic errors give rise to irregularly wavy wings on both sides of the central peak.

In Fig.~\ref{fig_520c_001_Fourier_total}, we show the effects of a $z$-polarized nuclear spin bath. Similarly, the colored dots denote the results simulated in a collaborative manner, where the ten
polarized nuclei with $\vec{p}^{(k)}=(0,0,1)$ are simulated on \texttt{ibm\_auckland} and the outer unpolarized nuclei are on \texttt{ibmq\_qasm\_simulator} according to the partition listed in
Table~\ref{tab_partition}, and the colored circles denote the counterpart fully performed on \texttt{ibmq\_qasm\_simulator}. As expected from the preliminary examinations, the collaborative simulations
on IBMQ for $\vec{p}^{(k)}=(0,0,1)$ are much better than those for $\vec{p}^{(k)}=(0,0,0)$ due to the corresponding polarization oracles. Moreover, the results fully given by
\texttt{ibmq\_qasm\_simulator} also fit the theoretical calculations very well besides the visible algorithmic errors in the imaginary parts. Furthermore, the profile of the CHER varies drastically with
increasing magnetic field in this case. Several sharp peaks emerge at strong fields. Remarkably, this phenomenon has also been well-reproduced in our simulations.

It has been shown that the nonclassicality is induced by the nuclear spin precession in the presence of a transversely polarized nuclear spin bath \cite{muche_process_n_cla_nvc_jpcm_2022}.
Figure~\ref{fig_520c_100_Fourier_total} shows the simulation of the nonclassicality induced by an $x$-polarized nuclear spin bath at various values of the magnetic fields. The polarization oracle
implementing $\vec{p}^{(k)}=(1,0,0)$ requires a quantum gate on the nucleus qubit to be polarized. After the amplification of the production over all $x$-polarized nucleus qubits on
\texttt{ibm\_auckland}, the errors in the collaborative simulations become prominent; while the overall profile remains visible. Similarly, the results fully given by
\texttt{ibmq\_qasm\_simulator} also suffer from the amplified algorithmic errors. Remarkably, in the lower panels of Fig.~\ref{fig_520c_100_Fourier_total}, we can observe the emergence of the
nonclassicality in terms of the negativity in the CHER $\wp(\omega)$. Although the nonclassicality is smeared at $B_z=50$ G due to the errors on \texttt{ibm\_auckland}, it becomes visible at stronger
fields, as shown in the insets for $B_z=100$ G and $200$ G.

\begin{figure*}[!ht]
\centering
\includegraphics[width=\textwidth]{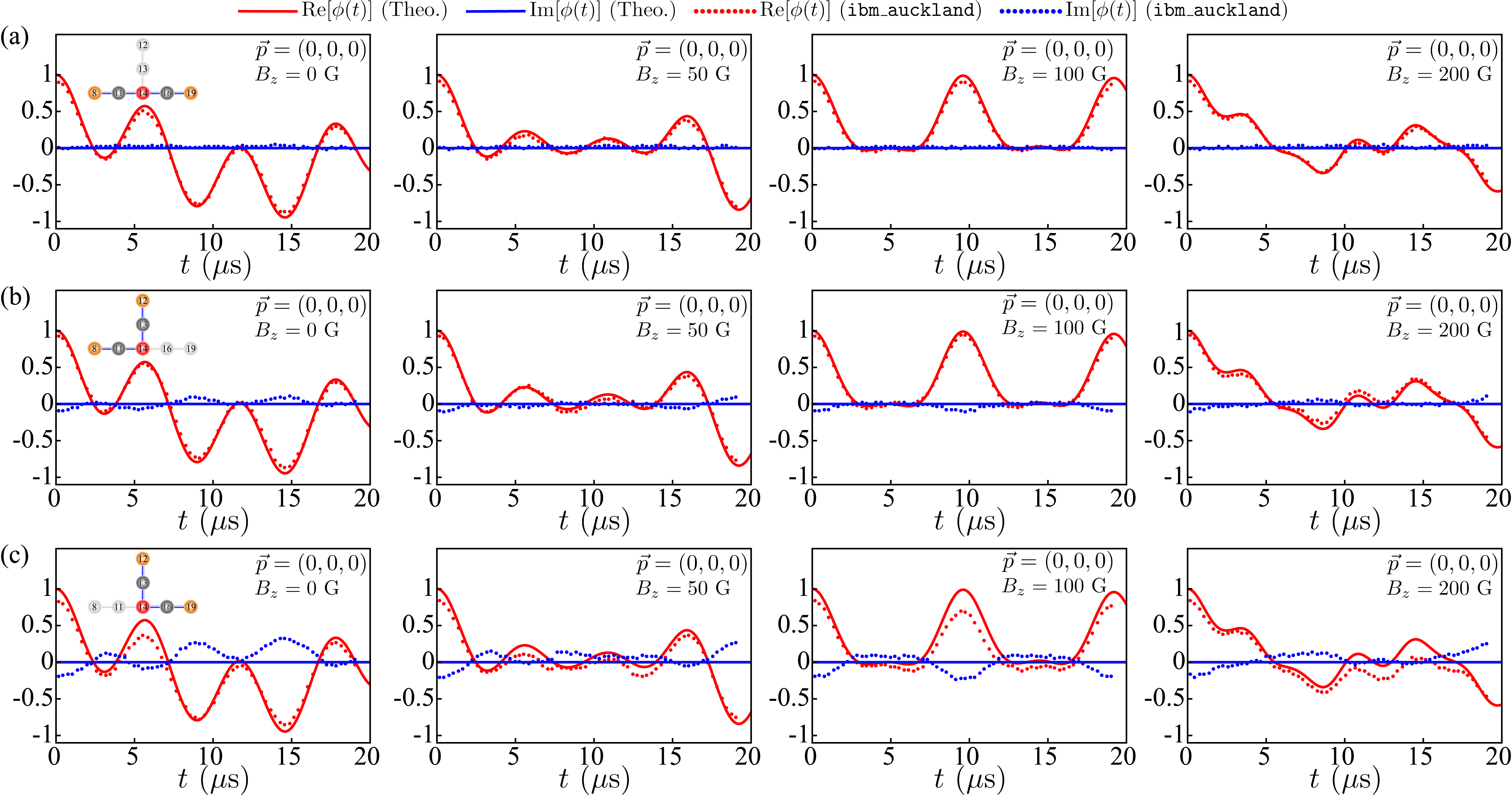}
\caption{The AQS results for two nuclei obtained from \texttt{ibm\_auckland} launching (a) the left-right, (b) the top-left, and (c) the top-right qubit pairs, respectively, as indicated in the
insets. The crosstalk is significantly suppressed by launching merely the left-right qubit pairs. The corresponding results are in good agreement with the theoretical calculations, particularly the null
imaginary part. However, the results given by the top-left and the top-right qubit pairs are subject to the noises caused by the crosstalk. The noises are prominent in the erroneous imaginary
parts.}
\label{fig_2c_total}
\end{figure*}

\begin{figure*}[!hb]
\centering
\includegraphics[width=\textwidth]{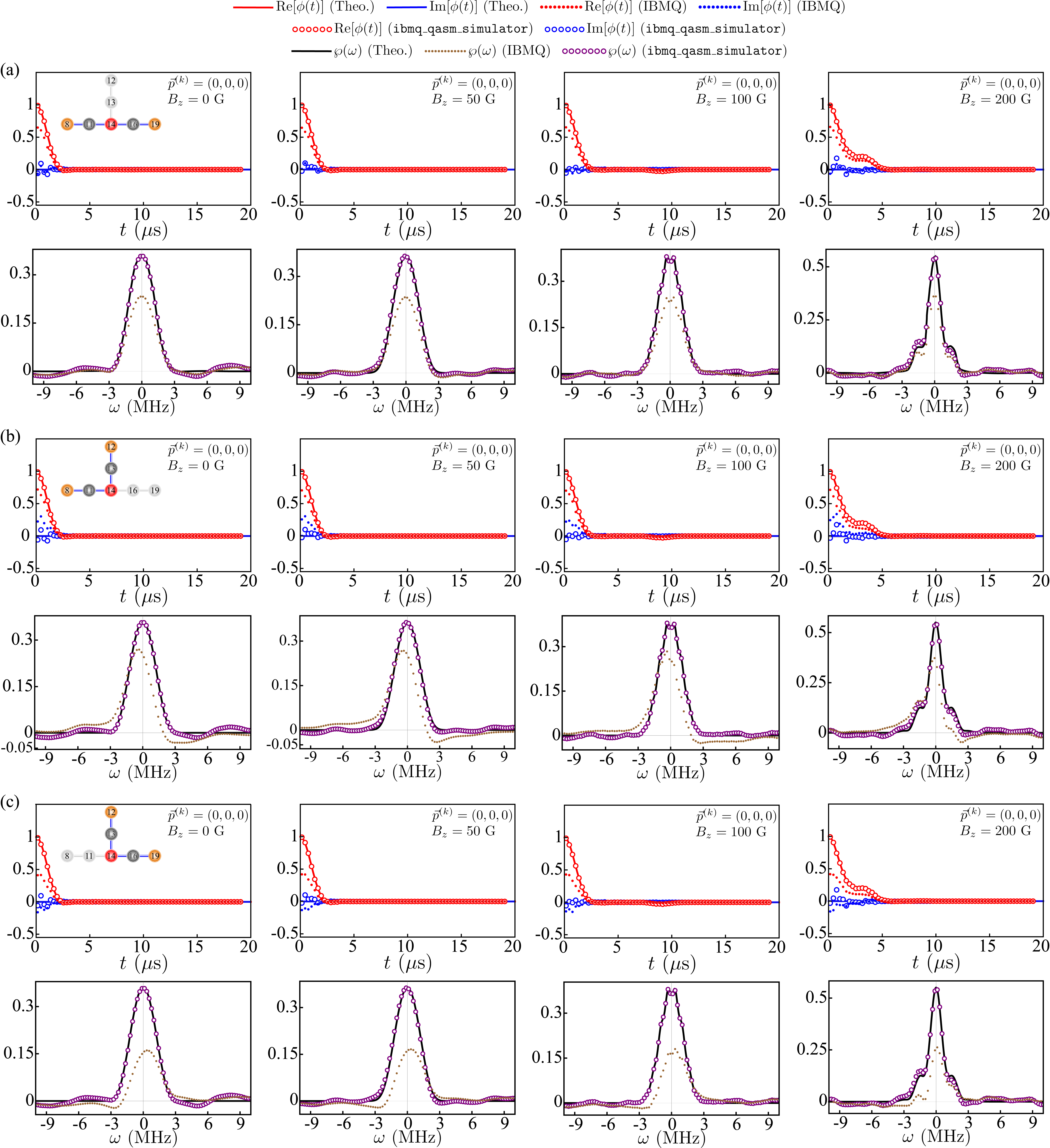}
\caption{The adaptively partitioned AQS results for 520 unpolarized nuclei at various values of the magnetic field and the corresponding CHER. The colored dots denote the results performed in a
collaborative manner with (a) the left-right, (b) the top-left, and (c) the top-right qubit pairs on \texttt{ibm\_auckland}, respectively, and \texttt{ibmq\_qasm\_simulator}. The results given by
the left-right qubit pairs exhibit significant improvements, leading to the elimination of the erroneous wings in the corresponding CHERs. However, the results given by the top-left or the top-right
qubit pairs are subject to the noises caused by the crosstalk, leading to prominent discrepancies in either real or imaginary parts, as well as the erroneous wings in the corresponding CHERs.
Additionally, the negative wings caused by the crosstalk indicate its nonclassical traits.}
\label{fig_520c_000_total_partition2c}
\end{figure*}

\section{NONCLASSICAL CROSSTALK BETWEEN QUBITS}

From the previous simulation results, it can be realized that, apart from the intrinsic local errors such as the gate errors or the finite life times of the qubits, our simulations are suffering
from an additional source of nonlocal noises, i.e., the crosstalk between qubits, which in turn constitute the primary obstacle hindering the numerical reliability of our simulations.

To address the nonlocal noises caused by the crosstalk, we perform the prototypical circuits simulating the effects of two $^{13}$C nuclei on \texttt{ibm\_auckland} launching different nucleus-ancilla
qubit pairs. Figure~\ref{fig_2c_total} shows the results for $\vec{p}^{(k)}=(0,0,0)$ at various values of the magnetic field. The insets indicate the qubits launched in the circuits.
It can be seen that, the circuits performed on the left-right qubit pairs [Fig.~\ref{fig_2c_total}(a)] significantly suppress the noises caused by the crosstalk. The results are in good agreement with
the theoretical calculations. Crucially, the results of null imaginary part $\mathrm{Im}[\phi(t)]$ have been correctly reproduced; while those performed on the top-right qubit pairs
[Fig.~\ref{fig_2c_total}(c)] give rise to the most prominent erroneous imaginary part. Additionally, the behavior of the errors are the same as those observed in the top panels of
Fig.~\ref{fig_3c_total}. These preliminary simulations not only confirm the effect of the crosstalk on the erroneous imaginary part, but also suggest a convenient way to suppress it by launching
appropriate qubits.

Based on these preliminary simulations, we apply our adaptive partition approach to simulate the effects of the 520 nuclei. Figure~\ref{fig_520c_000_total_partition2c} shows the results for
$\vec{p}^{(k)}=(0,0,0)$ at various values of the magnetic field launching different nucleus-ancilla qubit pairs, as indicated in the insets. Similarly, the colored dots denote the results performed in a
collaborative manner with \texttt{ibm\_auckland} and \texttt{ibmq\_qasm\_simulator}, and the colored circles denote the ones simulated with \texttt{ibmq\_qasm\_simulator}.
Compared with the FID process simulated in the top panels of Fig.~\ref{fig_520c_000_Fourier_total}, the results given by the left-right qubit pairs [Fig.~\ref{fig_520c_000_total_partition2c}(a)]
exhibit significant improvements.
The discrepancies in both real $\mathrm{Re}[\phi(t)]$ and imaginary parts $\mathrm{Im}[\phi(t)]$ from theoretical calculations are considerably quenched, as expected from Fig.~\ref{fig_2c_total}(a).
On the other hand, the results given by the top-left [Fig.~\ref{fig_520c_000_total_partition2c}(b)] or the top-right qubit pairs [Fig.~\ref{fig_520c_000_total_partition2c}(c)] are subject to the
noises caused by the crosstalk, leading to prominent discrepancies in either real or imaginary parts.

Noteworthily, the CHER $\wp(\omega)$ can further reveal different insights into the effect of the crosstalk.
Comparing the CHERs given by the left-right qubit pairs [Fig.~\ref{fig_520c_000_total_partition2c}(a)] with those shown in the bottom panels of Fig.~\ref{fig_520c_000_Fourier_total}, the erroneous
negative wings on the left, as well as the positive wings on the right, are eliminated due to the suppression of the crosstalk. Only the irregularly wavy wings caused by the algorithmic errors of
the simulator are left.

On the other hand, the CHERs given by the top-left [Fig.~\ref{fig_520c_000_total_partition2c}(b)] and the top-right qubit pairs [Fig.~\ref{fig_520c_000_total_partition2c}(c)] reveal erroneous
negative wings, indicating the nonclassical essential of the effect of the crosstalk.
It is also interesting to note that, the positions of the erroneous negative wings caused by different qubit pairs are different as well.
This implies that the relative phases between the wave functions of the spatially separated superconducting devices induced by the crosstalk are of different sign.

\section{CONCLUSION}

In this work we propose to simulate large-scale materials in a manner of analog quantum simulation on near-term quantum computing platforms. In view of the limitations on the computing capability
imposed by the noises and the topological connectivity, our simulation algorithm circumvents the obstacles by adaptively partitioning the effects of huge bath into adequate groups based on the
performance of the quantum devices.

We demonstrate our approach by simulating the FID process of the electron spin of an NV$^-$ center coupled to a huge nuclear spin bath and perform the simulation on IBMQ. We design a prototypical
quantum circuit implementing the total Hamiltonian of an NV$^-$ center coupled to a huge number of nuclei via the dipole-dipole hyperfine interaction. Additionally, to reflect the experimental
conditions, we also design a family of polarization oracles implementing the nuclear spin engineering by the DNP technique.

To investigate the capability of the quantum devices simulating the electron spin dynamics, we also perform a series of preliminary examinations simulating the effects of a few number of nuclei. Based on
their performance, we can simulate the FID process either in an collaboration with authentic device and classical simulator, or fully on classical simulator of IBMQ. With this adaptive partition
approach, we can reproduce the effects accounting for 520 nuclei on the FID process. In particular, we have taken into account the various values of magnetic fields and the nuclear spin polarizations in
an experimental condition. Additionally, by the technique of CHER, our approach can also reproduce the nonclassical essential of the electron spin FID process induced by the nuclear spin
polarizations.

Furthermore, we also notice that the simulation results are subject to imperfectness caused by both the noise of the authentic quantum devices and the algorithmic errors of the simulators. To further
showcase the versatility of our approach, we have also applied it to address the primary source of error in our simulations, i.e., the nonlocal noise caused by the crosstalk between qubits, and its
nonclassical essential. Our analyses suggest a convenient way to suppress it by launching appropriate qubits.

In conclusion, we achieve the demonstration of the capability of our adaptive partition approach in the exploration of the physical mechanisms underlying the simulated phenomena at a microscopic level.
Our approach reproduces critical physical phenomena, including the dynamical behavior of the electron spin, the variation of the profile of the CHER, the nonclassicality in terms of the negativity in the
CHER, and, crucially, the nonclassicality in the noises caused by the crosstalk between qubits. We stress that, our approach is flexible in the sense that we can distribute the computing loading not
only to different devices, but also to different qubit groups on a same device in a single task for improving the efficiency. Namely, the distribution strategy is adjustable depending on the condition
of the available devices and the required accuracy or efficiency.

($\equiv\widehat{\mathsf{\Phi}}\omega\widehat{\mathsf{\Phi}}\equiv$) $\sim$ meow

\section*{ACKNOWLEDGMENTS}

We acknowledge the NTU-IBM Q Hub and the IBM quantum experience for providing us a platform to implement the circuits. The views expressed are those of the authors and do not reflect the official
policy or position of IBM or the IBM quantum experience team.
This work is supported by the National Science and Technology Council, Taiwan, with Grants No. MOST 110-2112-M-006-012 and MOST 111-2112-M-006-015-MY3,
partially by the Higher Education Sprout Project, Ministry of Education to the Headquarters of University Advancement at NCKU,
and partially by the National Center for Theoretical Sciences, Taiwan.

\appendix 

\begin{figure*}[!ht]
\centering
\includegraphics[width=\textwidth]{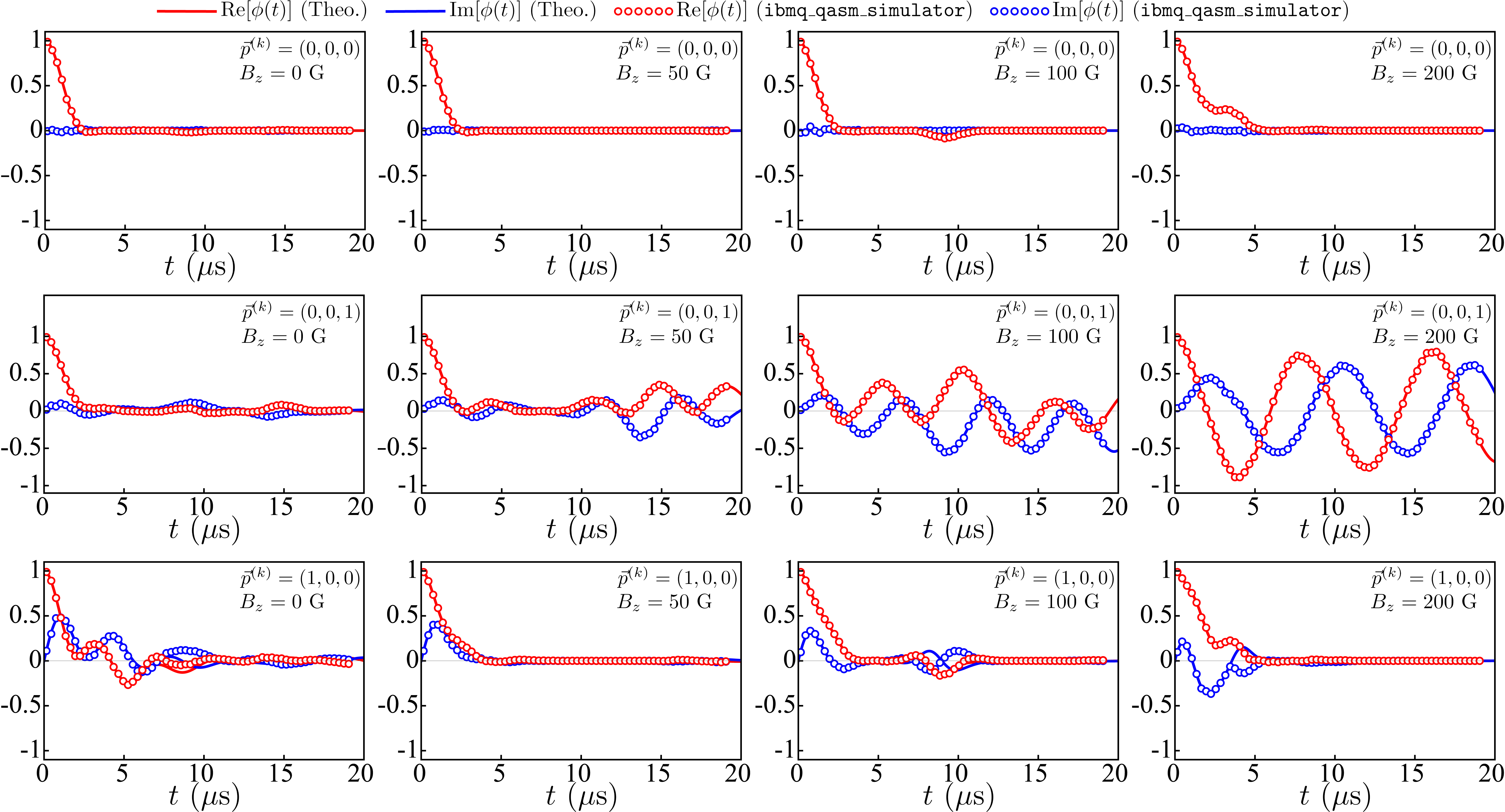}
\caption{The AQS results for ten nuclei obtained from \texttt{ibm\_qasm\_simulator}. We demonstrate the results of three polarizations, $\vec{p}^{(k)}=(0,0,0)$ (top panels), $\vec{p}^{(k)}=(0,0,1)$ (middle panels), and $\vec{p}^{(k)}=(1,0,0)$ (bottom panels), at various values of the magnetic field. We find that the classical simulator \texttt{ibm\_qasm\_simulator} can simulate at most ten
nucleus-ancilla qubit pairs in a single task. Regardless of the limitation on the number of qubits, the results fit the theoretical calculations very well besides tiny errors caused by the classical
simulation algorithm.}
\label{fig_10c_total}
\end{figure*}

\section{Parameters of the controlled-U gates in the AQS quantum circuit}\label{app_parameter_cu_aqs}

Here we explain how to determine the gate parameters $(\theta^{(k)},\varphi^{(k)},\lambda^{(k)},\gamma^{(k)})$ of the controlled-U gate in Eq.~(\ref{eq_total_unitary_qc_implementation}).

The matrix form of the U-gate to be controlled on IBMQ is expressed as
\begin{widetext}
\begin{equation}
e^{i\gamma^{(k)}}\widehat{U}(\theta^{(k)},\varphi^{(k)},\lambda^{(k)})
=e^{i\gamma^{(k)}}\left[\begin{array}{cc}
\cos\frac{\theta^{(k)}}{2} & -e^{i\lambda^{(k)}}\sin\frac{\theta^{(k)}}{2}\\
e^{i\varphi^{(k)}}\sin\frac{\theta^{(k)}}{2} & e^{i(\lambda^{(k)}+\varphi^{(k)})}\cos\frac{\theta^{(k)}}{2}
\end{array} \right].
\end{equation}
On the other hand, the relation to the total Hamiltonian is given by
\begin{eqnarray}
&&e^{i\gamma^{(k)}}\widehat{U}(\theta^{(k)},\varphi^{(k)},\lambda^{(k)})=\widehat{U}_1^{(k)}(t)\widehat{U}_0^{(k)\dagger}(t)
=\exp[-i(\vec{\Omega}_1^{(k)}\cdot\hat{\sigma}^{(k)})t/2]\times\exp[i(\Omega_0\hat{\sigma}^{(k)}_z)t/2] \nonumber\\
&&=\left[\begin{array}{cc}
\cos\frac{\Omega^{(k)}_1 t}{2}-i\sin\frac{\Omega^{(k)}_1 t}{2}u^{(k)}_z & -i\sin\frac{\Omega^{(k)}_1 t}{2}(u_x^{(k)}-iu_y^{(k)}) \\
-i\sin\frac{\Omega_1^{(k)}t}{2}(u_x^{(k)}+iu_y^{(k)}) & \cos\frac{\Omega_1^{(k)}t}{2}+i\sin\frac{\Omega_1^{(k)}t}{2}u_z^{(k)}
\end{array}\right]\left[\begin{array}{cc}
e^{i\frac{\Omega_0t}{2}} & 0 \\
0 & e^{-i\frac{\Omega_0t}{2}}
\end{array}\right] \nonumber\\
&&=e^{i\left(\frac{\Omega_0t}{2}+\Theta^{(k)}\right)}\left[\begin{array}{cc}
\sqrt{\cos^2\frac{\Omega_1^{(k)}t}{2}+\sin^2\frac{\Omega_1^{(k)}t}{2}u_z^{(k)2}} & -e^{i(\frac{\pi}{2}-\Omega_0t-\Theta^{(k)}-\Phi^{(k)})}\sin\frac{\Omega_1^{(k)}t}{2}\sqrt{u_x^{(k)2}+u_y^{(k)2}} \\
e^{i(-\frac{\pi}{2}-\Theta^{(k)}+\Phi^{(k)})}\sin\frac{\Omega_1^{(k)}t}{2}\sqrt{u_x^{(k)2}+u_y^{(k)2}} & e^{i\left(-\Omega_0t-2\Theta^{(k)}\right)}\sqrt{\cos^2\frac{\Omega_1^{(k)}t}{2}+\sin^2\frac{\Omega_1^{(k)}t}{2}u_z^{(k)2}}
\end{array}\right],
\end{eqnarray}
where $\Omega_0=\gamma_\mathrm{C}B_z$, $\vec{\Omega}_1^{(k)}=\vec{A}_z^{(k)}+\vec{\Omega}_0$, $\vec{u}^{(k)}=\vec{\Omega}_{1}^{(k)}/|\vec{\Omega}_{1}^{(k)}|$, $\Theta^{(k)}=\mathrm{Arg}[\cos\frac{\Omega_1^{(k)}t}{2}-i\sin\frac{\Omega_1^{(k)}t}{2}u_z^{(k)}]$, and $\Phi^{(k)}=\mathrm{Arg}[u_x^{(k)}+iu_y^{(k)}]$.
By comparing the above equations, we obtain the results shown in Eq.~(\ref{eq_parameter_cu_aqs}).
\end{widetext}

\section{Simulating ten nuclei on a simulator}\label{simulate 10 nuclei on simulator}

Limited by the topological connectivity, it is infeasible to simulate huge materials in a single task on the IBMQ authentic devices. To circumvent this limitation, as well as to benchmark the capability
of the classical simulators provided by IBMQ, we have also performed larger prototypical circuits on the \texttt{ibm\_qasm\_simulator}.

Although the \texttt{ibm\_qasm\_simulator} provides 32 qubits, we find that it has a limited computing capability simulating up to ten nucleus-ancilla qubit pairs (21 qubits launched in a single task).
Errors occur in the backend operation if more than 21 qubits are included in a single task. This limitation can be understood from the giant Hilbert space of size $2^{21}$, corresponding to the
propagation of a density matrix of dimension $2^{21}\times 2^{21}$.

In Fig.~\ref{fig_10c_total}, we show the prototypical simulations for the effects of ten $^{13}$C nuclei on \texttt{ibm\_qasm\_simulator}. It can be seen that the results given by the simulator fit
the theoretical calculations very well for three polarizations. However, there are still tiny errors due to the approximations introduced by classical simulation algorithm.


%

\end{document}